\documentclass[12pt]{article}

\ifx\pdfoutput\undefined
\usepackage[dvips,bookmarks]{hyperref}
\else
\usepackage{hyperref}
\fi
\hypersetup{colorlinks=false,bookmarksopen,bookmarksnumbered,citecolor=blue,
   pdfstartview=FitH}

\usepackage{latexsym}
\usepackage{amssymb,amsfonts,amsmath}
\usepackage{graphicx} 
\usepackage{indentfirst}
\usepackage{bbm}
\usepackage{amssymb}
\usepackage{verbatim}
\usepackage{amsmath, amsthm,amssymb}
\usepackage{mathrsfs}
\usepackage{hyperref}
\usepackage{amsfonts}
\usepackage{dsfont}
\usepackage{slashed, tensor}
\usepackage{booktabs}
\usepackage{graphicx}
\usepackage{mathrsfs}
\usepackage[nosort]{cite}

\parskip=\medskipamount

\newcommand {\cA}{{\cal A}}
\newcommand {\cB}{{\cal B}}
\newcommand {\cC}{{\cal C}}
\newcommand {\cD}{{\cal D}}
\newcommand {\cE}{{\cal E}}

\newcommand {\cH}{{\cal H}}

\newcommand {\cJ}{{\cal J}}
\newcommand {\cK}{{\cal K}}
\newcommand {\cL}{{\cal L}}
\newcommand {\cM}{{\cal M}}
\newcommand {\cN}{{\cal N}}
\newcommand {\cO}{{\cal O}}

\newcommand {\cR}{{\cal R}}

\newcommand {\cT}{{\cal T}}

\newcommand {\cW}{{\cal W}}

\def\a{\alpha}
\def\b{\beta}

\def\d{\delta}
\def\e{\epsilon}

\def\g{\gamma}
\def\G{\Gamma}

\def\l{\lambda}

\def\o{\omega}

\def\q{\theta}
\def\r{\rho}
\def\s{\sigma}
\def\t{\tau}

\def\x{\xi}

\def\D{\Delta}
\def\F{\Phi}

\def\L{\Lambda}

\def\ri{{\rm i}}
\def\re{{\rm e}}

\newcommand{\ve}{\varepsilon}

\newcommand{\pa}{\partial}                
\newcommand{\hf}{\frac12}

\newcommand{\be}{\begin{equation}}
\newcommand{\ee}{\end{equation}}
\newcommand{\bea}{\begin{eqnarray}}
\newcommand{\eea}{\end{eqnarray}}
\newcommand{\non}{\nonumber}
\newcommand{\ba}{\begin{array}}
\newcommand{\ea}{\end{array}}

\newcommand{\1}{{\underline{1}}}
\newcommand{\2}{{\underline{2}}}

\newcommand{\bm}[1]{\mbox{\boldmath$#1$}}

\def\double #1{#1{\hbox{\kern-2pt $#1$}}}

\newcommand{\ha}{{\hat{a}}}
\newcommand{\hb}{{\hat{b}}}

\newcommand{\bbD}{{\mathbb {D}}}

\newcommand{\sSU}{\mathsf{SU}}

\newcommand{\sSO}{\mathsf{SO}}
\newcommand{\sU}{\mathsf{U}}

\newcommand{\bsubeq}{\begin{subequations}}
\newcommand{\esubeq}{\end{subequations}}

\newcommand{\eps}{{\ve}}

\newcommand{\dbeta}{{\dot{\beta}}}
\newcommand{\dgamma}{{\dot{\gamma}}}
\newcommand{\ddelta}{{\dot{\delta}}}

\newcommand{\eol}{\notag \\}
\newcommand{\rd}{\mathrm d}
\newcommand{\veps}{\varepsilon}

\numberwithin{equation}{section}

\newcommand{\RM}{R(M)}
\newcommand{\RD}{R(\mathbb D)}

\newcommand{\RK}{R(K)}
\newcommand{\sRM}{R(M)}
\newcommand{\sRD}{R(\mathbb D)}

\newcommand{\sRJ}{R(J)}
\newcommand{\sRS}{R(S)}
\newcommand{\sRK}{R(K)}
\newcommand{\scT}{{T}}

\DeclareMathOperator{\Tr}{Tr}

\begin{document}

\begin{titlepage}
\begin{flushright}
Nikhef-2016-026\\
June, 2016 \\
\end{flushright}
\vspace{5mm}

\begin{center}
{\Large \bf Invariants for minimal conformal supergravity \\
in six dimensions
}
\\ 
\end{center}

\begin{center}

{\bf
Daniel Butter${}^{a}$, Sergei M. Kuzenko${}^{b}$, Joseph Novak${}^{c}$ \\
and
Stefan Theisen${}^{c}$
} \\
\vspace{5mm}

\footnotesize{
${}^{a}${\it Nikhef Theory Group \\
Science Park 105, 1098 XG Amsterdam, The Netherlands}}
~\\
\texttt{dbutter@nikhef.nl}\\
\vspace{2mm}

\footnotesize{
${}^{b}${\it School of Physics M013, The University of Western Australia\\
35 Stirling Highway, Crawley W.A. 6009, Australia}}  
~\\
\texttt{sergei.kuzenko@uwa.edu.au}\\
\vspace{2mm}

\footnotesize{
${}^{c}${\it Max-Planck-Institut f\"ur Gravitationsphysik, Albert-Einstein-Institut,\\
Am M\"uhlenberg 1, D-14476 Golm, Germany.}
}\\
\texttt{joseph.novak,\,stefan.theisen@aei.mpg.de}\\
\vspace{2mm}

\end{center}

\begin{abstract}
\baselineskip=14pt
We develop a new off-shell formulation for six-dimensional conformal supergravity obtained by gauging  the 
6D $\cN = (1, 0)$ superconformal algebra in superspace. 
This formulation 
is employed 
to construct two 
invariants for 6D $\cN = (1, 0)$ conformal supergravity, 
which contain $C^3$ and $C\Box C$ terms at the component level.
Using a conformal supercurrent analysis, we prove 
that these exhaust all such invariants 
in 
minimal
conformal supergravity. 
Finally, we show how to construct the supersymmetric 
$F \Box F$ invariant in curved superspace.
\end{abstract}

\vfill

\vfill
\end{titlepage}

\newpage
\renewcommand{\thefootnote}{\arabic{footnote}}
\setcounter{footnote}{0}

\tableofcontents


\allowdisplaybreaks

\section{Introduction}

Conformal field theories (CFTs) play a distinguished role among relativistic 
quantum field theories. It has long been realized that they arise 
as fixed point theories of renormalization group flows and  
the study of their properties is clearly of interest. The enlarged symmetry 
group helps to constrain e.g. the general structure of correlation functions
beyond what is already required by Poincar\'e invariance. 
Additional symmetries lead to further restrictions. One such symmetry which 
is very powerful in this respect is supersymmetry, in which case one deals 
with superconformal field theories (SCFTs). 

It has been known since the early days of supersymmetry that superconformal 
theories can only exist in six or lower 
dimensions\cite{Nahm}.
In six dimensions, where $\cN=(p,q)$ Poincar\'e superalgebras
exist for any integer $p,q\geq 0$, superconformal algebras  
only exist for either $p=0$ or $q=0$. In fact, the 
only known non-trivial unitary CFTs in six dimensions are supersymmetric and 
arise as world-volume theories
of appropriate brane configurations in string and M-theory and in F-theory, in the 
limit where gravity decouples.
They realize either $\cN=(2,0)$ or $\cN=(1,0)$ superconformal symmetry. 
For these theories no Lagrangian description is known but 
they are believed to obey the axioms of quantum field theories.\footnote{Here we are concerned 
with unitary SCFTs. For an example of a higher-derivative classical SCFT, see \cite{ISZ05}.} 
They should, in particular, have local conserved current operators and among them  
a local conserved and traceless energy-momentum tensor \cite{SW,Seiberg}.   
Evidence for the existence of $\cN=(2,0)$ theories was first given in 
\cite{Witten1,Strominger,Witten2}; for $\cN=(1,0)$ theories we refer to 
\cite{SW,Seiberg,GH,BI,HZ,HMV,DZHTV,HMRV}. 

As mentioned before, symmetries in quantum field theories lead to restrictions 
on correlation functions which have to satisfy Ward identities. 
In correlation functions of conserved currents one finds, however, that 
the naive Ward identities which would follow from the symmetries cannot 
always be satisfied simultaneously. This happens in even dimensions and leads to 
(super)conformal anomalies which express the fact that imposing conservation and 
tracelessness of the energy-momentum tensor clashes in certain 
correlation functions. The general structure of these conformal or Weyl
anomalies was analyzed by Deser and Schwimmer \cite{DS} who also introduced the 
classification into two types: type A and type B. 
In any even dimension there is always one type A anomaly and 
starting in four dimensions, an increasing number of type B 
anomalies. The easiest way to discuss them 
is to couple the conformal field theory to a metric background which serves as a 
source for the energy-momentum tensor. The anomalies then express the 
non-invariance of the effective action (generating functional) under 
a local Weyl rescaling of the metric. 
The anomalous variation of the non-local effective action results in 
anomalies which are local diffeomorphism invariant functions of the metric and 
its derivative, i.e. functions of the curvature and its covariant derivatives. 
The type A anomaly in any even dimension is given by the Euler density of 
that dimension; the type B anomalies are Weyl invariant expressions constructed 
from the curvature tensors and its covariant derivatives\cite{DS}. In four dimensions there 
is one such expression, the square of the Weyl tensor; in six dimensions 
there are two inequivalent contractions of three Weyl tensors and one Weyl invariant 
expression which involves two covariant derivatives. If we work in a 
topologically trivial background, only the type B anomalies contribute 
if one rescales the metric by a constant factor. 

In any dimension the possible Weyl anomalies can be found by imposing the Wess-Zumino 
consistency condition\cite{WZ}, 
which expresses the obvious fact that two consecutive Weyl 
variations of the effective action must commute. 
Non-supersymmetric CFTs are then characterized by as many 
anomaly coefficients as there are solutions to the Wess-Zumino consistency condition:
one in two, two in four and four in six dimensions, respectively. 

In SCFTs, the Weyl anomalies are accompanied by superconformal 
and $R$-symmet\-ry anomalies; 
altogether they constitute the so-called 
super-Weyl anomalies. 
They are related by supersymmetry and 
various anomalies in bosonic and fermionic symmetry currents  
are packaged into anomaly supermultiplets.  
The most elegant way to exhibit this is using  
a mani\-festly supersymmetric formulation, i.e. superspace.  
In four dimensions, the super-Weyl anomalies were studied in \cite{BPT,BK86}
in the $\cN=1$ case and in \cite{K_N=2SW} for $\cN=2$. 
Furthermore,  supersymmetry might also reduce the number of independent anomaly 
coefficients by packaging several solutions of the Wess-Zumino consistency 
conditions into one supermultiplet. This is the case for ${\cal N}=4$ 
supersymmetric Yang-Mills theory in four dimensions where 
there is only one independent anomaly coefficient.

As Lagrangian descriptions of six-dimensional SCFTs are not known, 
it is rather difficult to study their dynamics. Interesting non-trivial information  
can, however, be obtained from their symmetries. One can e.g. show that
$\cN=(2,0)$ and $\cN=(1,0)$ SCFTs have neither marginal nor relevant 
supersymmetry preserving deformations \cite{CDI2,LL}. Another way to 
approach these theories is via their 't Hooft and Weyl 
anomalies. This was done in 
\cite{FHMM,HS,HMM,Intriligator,Yi,OSTY,CDY}.  

Due to supersymmetry one expects that the two types of anomalies are 
para\-metrized by the same coefficients. This   
is known e.g. for ${\cal N}=1$ SCFT in four dimensions, where the 
U(1) $R$-current anomalies are governed by linear combinations of the two 
independent Weyl anomaly coefficients.  It would be useful to know similar relations 
for SCFTs in six dimensions and furthermore, to know the precise number 
of independent anomaly coefficients. We consider the analysis of this paper as a first 
step towards answering these questions for $\cN= (1,0)$ SCFTs. More precisely, we will 
construct supersymmetry invariants which contain the solutions of the 
WZ consistency condition for the Weyl anomaly as one of their bosonic components. 
By supersymmetry, these 
invariants should contain the solutions to the supersymmetrized version 
of the WZ condition. 
Here we content ourselves with the first step, the 
construction of the supersymmetric invariants and leave 
a detailed analysis of the anomaly structure for the future. 
But the results of this paper already show that the number of anomaly coefficients 
is reduced: while in the non-supersymmetric case there are three independent  
type B Weyl anomalies, i.e. dimension six combinations 
of curvature tensors and covariant derivatives which transform 
homogeneously under Weyl transformations of the metric,  
there are only two independent superspace invariants which contain them. 
In addition to their relevance for the anomaly structure, 
their arbitrary linear combination is the action for minimal conformal supergravity 
in six dimensions, which 
will be the main focus of this paper.

To establish these results we develop a new off-shell superspace formulation 
of this theory. 
We therefore start with a brief review of  
six-dimensional (6D) minimal conformal supergravity and conformal superspace methods (see \cite{FT85} for a review of 
conformal supergravity theories in 4D). 
Its superconformal tensor calculus 
was formulated thirty years ago by Bergshoeff, Sezgin and 
Van Proeyen \cite{BSVanP}. In many respects, 
it is analogous to the superconformal tensor calculus for  4D $\cN=2$ supergravity
\cite{deWvHVP1,deRvHdeWVP,deWvHVP2,deWPVP,deWLPSVP,deWLVP}, 
see \cite{FVP} for a recent pedagogical review.
Soon after 
the 6D $\cN=(1,0)$ superconformal method \cite{BSVanP} appeared, it
was applied to construct the off-shell 
supersymmetric extension
of the Riemann curvature squared term  \cite{BSS1,BSS2,BR}.
More recently, the 6D $\cN=(1,0)$ superconformal techniques of \cite{BSVanP} 
have been refined  \cite{CVanP,BCSVanP}. In particular, the complete 
off-shell action for minimal Poincar\'e supergravity has been given in \cite{CVanP}  
(only the bosonic part of this action was explicitly worked out in \cite{BSVanP}). 
Gauged minimal 6D supergravity 
has been worked out in \cite{BCSVanP} by coupling the minimal supergravity
of \cite{CVanP}  
to an off-shell vector multiplet.    
The resulting theory  is an off-shell version of the dual formulation \cite{NS2,FRS}
of the Salam-Sezgin model \cite{SalamS,NS1}. 

 Similar to the 4D $\cN=2$  case, 
the 6D $\cN=(1,0)$ superconformal tensor calculus 
has two limitations. Firstly, it does not provide tools to describe off-shell hypermultiplets.
Only on-shell hypermultiplets were used in \cite{BSVanP}
as well as in all later developments based on \cite{BSVanP}.
Secondly, it does not offer insight  as to how general  
higher-derivative supergravity actions can be 
built, see \cite{VanP} for a recent discussion. 
In particular, (off-shell) invariants for 6D $\cN=(1,0)$ conformal supergravity have 
never been constructed. In order to avoid these limitations, 
one has to resort to superspace techniques. 
At this point, some comments are in order about the superspace approaches 
to conformal supergravity in diverse dimensions.

There are two general approaches to describe $\cN$-extended 
conformal supergravity in $D \leq 6$ dimensions\footnote{The cases
of five and six dimensions are rather special. 
Conformal supergravity exists only for $\cN=1$ in five dimensions, 
and only for  $\cN=(p,0)$ in six dimensions.}
using a curved $\cN$-extended superspace $\cM^{D|\d}$, where 
$\d$ denotes the number of fermionic dimensions.  
One of them, known as  $G_R^{[D;\cN]}$ superspace, 
 makes use of the superspace structure group
$\sSO(D-1,1) \times G_R^{[D;\cN]}$, 
where $\sSO(D-1,1)$ is the Lorentz group 
and $G_R^{[D;\cN]}$ is the $R$-symmetry group of the $\cN$-extended 
super-Poincar\'e algebra in $D$ dimensions.\footnote{The group $G_R^{[D;\cN]}$ coincides with $\sSO(\cN)$ for $D=3$, 
$\sU(\cN)$ for $D=4$, 
and $\sSU(2)$ for the cases 5D $\cN=1$ and 6D $\cN=(1,0)$.} 
A fundamental requirement on the superspace geometry, 
which should describe conformal supergravity,  
is that the constraints on the superspace torsion be invariant under 
a super-Weyl transformation generated by a real unconstrained superfield 
parameter. This approach was pioneered in four dimensions by 
 Howe \cite{Howe1,Howe2} who fully developed the $\sU(1)$ and $\sU(2)$ 
 superspace geometries  \cite{Howe2} corresponding to 
 the  $\cN=1$ and $\cN=2$ cases, respectively.  
 The superspace formulation for 5D conformal 
supergravity (5D $\sSU(2)$ superspace)
was presented in \cite{KT-M08}, 
and it
was naturally extended to the 6D $\cN=(1,0)$ case in \cite{LT-M12}
where 6D $\sSU(2)$ superspace was formulated.
In three dimensions, the $\sSO(\cN)$ superspace 
geometry was developed in \cite{HIPT,KLT-M}. 

The other superspace approach to conformal supergravity is based on gauging the entire 
$\cN$-extended superconformal group in $D$ dimensions,
of which $\sSO(D-1,1) \times G_R^{[D;\cN]}$ is a subgroup. 
This approach, known as conformal 
superspace, was originally developed
for $\cN=1$  and $\cN=2$ supergravity theories in four dimensions
by one of us (DB) \cite{Butter4DN=1,Butter4DN=2}.
More recently, it has been extended to the cases of 
3D $\cN$-extended conformal 
supergravity  \cite{BKNT-M1} and  5D conformal supergravity
\cite{BKNT-M5D}. Conformal superspace is a more general formulation 
than $G_R^{[D;\cN]}$ superspace in the sense that the latter 
is obtained from the former by partially fixing the gauge freedom, 
see \cite{Butter4DN=1,Butter4DN=2,BKNT-M1,BKNT-M5D} for more details.

Unlike the superconformal tensor calculus, the superspace method offers
off-shell formulations for the most general supergravity-matter couplings
with eight supercharges in four, five and six dimensions.
This includes off-shell formulations for hypermultiplets and 
their most general locally supersymmetric sigma model couplings.
The first such formulations were developed using harmonic superspace
\cite{Galperin:1985zv, Galperin:1987ek, Bagger:1987rc}
(see also \cite{GIOS}) and employed explicit supergravity prepotentials
(but see \cite{Galperin:1987em, Butter:2015nza} for covariant approaches).
Later, off-shell geometric formulations were derived for
5D $\cN=1$ supergravity-matter systems  \cite{KT-M_5D3,KT-M08} 
by putting forward
the novel concept of covariant projective multiplets.
These supermultiplets are curved-superspace extensions of  
the 4D $\cN=2$ and 5D $\cN=1$ superconformal projective multiplets \cite{K06,K07}. 
The latter reduce to the off-shell projective multiplets 
pioneered by Lindstr\"om and Ro\v{c}ek 
\cite{KLR,LR1,LR2} in the 4D $\cN=2$ super-Poincar\'e case.
The 5D off-shell formulations  have been  generalized to the 4D $\cN=2$ 
 \cite{KLRT-M_4D-1,KLRT-M_4D-2}, 3D $\cN=4$ \cite{KLT-M}
 and 6D $\cN=(1,0)$ \cite{LT-M12}
 cases.\footnote{In the 6D $\cN=(1,0)$ super-Poincar\'e case, the projective-superspace
 formalism was introduced in \cite{GL}, where it was used to construct
 off-shell actions for self-interacting linear multiplets.}
 All of these works made use of 
 the appropriate $G_R^{[D;\cN]}$ superspace. However, 
 all the results are naturally lifted to conformal
superspace.

Conformal superspace is an ideal setting to reduce 
the locally supersymmetric actions from superspace to components 
\cite{BN12,Butter-hyper}.
It also turns out to be an efficient formalism to build general  
higher-derivative supergravity actions. 
Recent applications of the conformal superspace approach have involved
constructing
(i) the $\cN$-extended conformal supergravity 
actions in three dimensions for $3\leq \cN \leq 6$ \cite{BKNT-M2,KNT-M}, and 
(ii) new higher-derivative invariants in 4D $\cN=2$ supergravity, 
including the Gauss-Bonnet term \cite{BdeWKL}.
In the present paper, we develop 6D $\cN=(1,0)$
conformal superspace and apply it to construct invariants 
for conformal supergravity. 

Before turning to the details of the six-dimensional case, it is worth recalling 
the structure of 
conformal supergravity actions in four dimensions
(see for example the reviews \cite{Muller, FT85}). 
The invariants for $\cN < 3$ are supersymmetric extensions of the $C^2$ term 
and are described by chiral integrals of the form
\bea 
I_{C^2} := \int \rd^4 x \, \rd^{2 \cN} \q \, \cE \, 
W^{\a_1 \dots \a_{4-\cN}}W_{\a_1 \dots \a_{4-\cN}}
+ {\rm c.c.} \ , 
\qquad \cN = 1, 2,
\eea
where $\cE$ is the chiral integration measure. 
The covariantly chiral tensor superfield
$W_{\a_1 \dots \a_{4-\cN}} = W_{(\a_1 \dots \a_{4-\cN})}$
 is the 
superspace generalization
of the Weyl tensor (known as the  super-Weyl tensor). 
Thus the structure of 4D $\cN$-extended conformal supergravity is 
remarkably simple for $\cN<3$. 

The case of 6D $\cN=(1,0)$ conformal supergravity has conceptual differences from its 4D $\cN=2$ cousin. First of all, there is no covariantly defined chiral subspace 
of SU(2) superspace \cite{LT-M12}, and thus we cannot generalise
the 4D $\cN=2$ construction to six dimensions.  Of course, one could try and construct 
invariants for conformal supergravity as full superspace integrals of the form 
\begin{align}
S = \int \rd^6x\, \rd^8\q\, E\, \cL \ ,
\end{align}
where the Lagrangian $\cL$ is a real primary superfield of dimension 2
(in the sense of \cite{LT-M12}). This Lagrangian should 
be constructed in terms of the dimension-1 super-Weyl tensor $W^{\a\b} = W^{ \b \a}$ 
 \cite{LT-M12} and its covariant derivatives. 
It is obvious that no  $\cL$ with the required properties exists. 
In the case of 4D $\cN=2$ supergravity, it was shown \cite{K-08,KT-M-09}
that the chiral action principle can be reformulated as a special case 
of the 4D $\cN=2$ projective-superspace action \cite{KLRT-M_4D-1,KLRT-M_4D-2}. 
For supergravity theories with eight supercharges
in diverse dimensions (including the 3D $\cN=4$ \cite{KLT-M}, 
5D $\cN=1$ \cite{KT-M08} and 6D $\cN=(1,0)$ \cite{LT-M12} cases),
the projective-superspace action principle is known to be universal in the 
sense that it can be used to realize general off-shell supergravity-matter couplings. 
The same statement holds for harmonic superspace (see \cite{Sokatchev:PhiPrepot} for the
6D $\cN=(1,0)$ case in particular). If the goal were to build 
two-derivative supergravity-matter actions, either approach would
suffice. However, if one is interested in realizing
the invariants for 6D $\cN=(1,0)$ conformal supergravity, 
it proves to be impossible to construct any  projective-superspace 
Lagrangian $\cL^{ (2)}$  only in terms  of the super-Weyl tensor $W^{\a\b}$,
that is without introducing prepotentials for the Weyl multiplet;
and while 6D harmonic superspace furnishes such explicit prepotentials, the problem of
constructing the necessary higher derivative invariants (while respecting the prepotential
gauge transformations) remains a challenge unsolved even in the 4D case, where
the covariant actions are known.
Therefore, if one is interested in constructing the invariants for 6D $\cN = (1,0)$ conformal supergravity
solely in terms of the covariant super-Weyl tensor, a new action principle is required. 
The present paper addresses this problem and 
demonstrates that there are two action principles which naturally support all the 6D $\cN = (1, 0)$ Weyl invariants.

This paper is organized as follows. Section \ref{confGravity} is a review on
conformal gravity and includes a simple derivation of the 6D Weyl invariants. 
In section \ref{setup} we describe 
6D $\cN = (1, 0)$ conformal superspace. 
In section \ref{actPrinc1} an action principle is presented in conformal superspace and it is 
shown how it can be used to describe a supersymmetric invariant containing a $C^3$ term. Application 
to other invariants is also discussed. Section \ref{actPrinc2} is devoted to deriving another action principle 
which is used to describe a supersymmetric invariant containing a $C \Box C$ term 
and a higher derivative action based on the Yang-Mills multiplet in conformal superspace. Concluding comments and a discussion are 
given in section \ref{conclusion}, where it is proved that the 6D $\cN = (1, 0)$ Weyl invariants constructed 
exhaust all such invariants in minimal conformal supergravity.

We have included a number of technical appendices. In Appendix \ref{NC} we include a summary of 
our notation and conventions. Appendix \ref{KVF} is devoted to a derivation of the superconformal algebra from the algebra 
of conformal Killing supervector fields of 6D $\cN = (1,0)$ Minkowski superspace. Finally, in Appendix \ref{YMmultiplet} 
we give a description of the Yang-Mills multiplet in conformal superspace.


\section{Conformal gravity in six dimensions}
\label{confGravity}

The conformal invariants in six dimensions \cite{BPB,DS,KMM} 
have been constructed previously and are well known. Since 
we will be concerned with their supersymmetric generalizations, it is 
natural to first present their bosonic counterparts. In this section, we provide 
a simple derivation of the 
conformal invariants. 
The formulation we use here will be naturally generalized 
to the supersymmetric case in later sections and will serve 
as a prelude to the conformal superspace formulation in section \ref{setup}.
We begin by reviewing the formulation for conformal gravity 
in $D >3$ spacetime dimensions following 
\cite{BKNT-M1}.\footnote{Conformal gravity 
has been discussed elsewhere in many places, e.g. \cite{FVP}. Our review
here emphasizes certain points relevant to our paper.}


\subsection{Conformal gravity in $D>3$  spacetime dimensions}

The conformal algebra in $D>2$ spacetime dimensions, $\frak{so}(D, 2)$, 
is spanned by the generators $X_{\underline{a}} = \{ P_a, M_{ab}, \mathbb D , K_a\}$, 
which obey the commutation relations 
\begin{subequations}
\begin{gather}
[M_{ab} , M_{cd}] = 2 \eta_{c[a} M_{b] d} - 2 \eta_{d [a} M_{b] c} \ , \\
[M_{ab} , P_c ] = 2 \eta_{c [a} P_{b]} \ , \quad [\mathbb D, P_a] = P_a \ , \\
[M_{ab} , K_c] = 2 \eta_{c[a} K_{b]} \ , \quad  [\mathbb D, K_a] = - K_a \ , \\
[K_a , P_b] = 2 \eta_{ab} \mathbb D + 2 M_{ab} \ ,
\end{gather}
\end{subequations}
where $P_a$ is the translation, $M_{ab} = - M_{ba}$ is the Lorentz, $\bbD$ is the dilatation and 
$K_a$ is the special conformal generator.

To describe conformal gravity one begins with a $D$-dimensional manifold $\cM^D$ 
parametrized by local coordinates $x^m$, $m = 0, 1, \cdots , D - 1$. Following the gauging procedure 
in \cite{BKNT-M1}, the covariant derivatives are chosen to have the form
\be
\nabla_a =  e_a - \hf \omega_a{}^{bc} M_{bc} - b_a \mathbb D - \mathfrak{f}_a{}^b K_b \ .
\ee
Here $e_a = e_a{}^m \partial_m$ is the inverse vielbein, 
while $\omega_a{}^{bc}$ is the Lorentz, $b_a$ is the dilation and $\frak{f}_a{}^b$ is the special conformal 
connection, respectively. The covariant derivatives may also be cast in the framework of forms
\be \nabla = e^a \nabla_a = \rd - \hf \omega^{bc} M_{bc} - b \, \mathbb D - \mathfrak{f}^a K_a \ ,
\ee
where $e^a := \rd x^m e_m{}^a$ is the vielbein, $\rd$ is the exterior derivative and we have defined
$\omega^{bc} := e^a \omega_a{}^{bc}$,  $b := e^a b_a$ and
$\mathfrak{f}^a := e^b \mathfrak{f}_b{}^a$.

The gravity gauge group is generated by local transformations 
which can be summarised by\footnote{One must take care in 
applying this formula since one can have $\L^{\underline{a}} = 0$ but $\nabla_a \L^{\underline{b}} \neq 0$.}
\be \d_{\cK} \nabla_a = [\cK , \nabla_a] \ , \quad \cK =  \xi^a \nabla_a + \L^{\underline{a}} X_{\underline{a}} 
=  \xi^a \nabla_a 
+ \hf \L(M)^{ab} M_{ab} + \s \bbD + \L(K)^a K_a
\ee
provided we interpret
\be
\nabla_a \xi^b := e_a \xi^b + \omega_a{}^{\underline{c}} \xi^d f_{d \underline{c}}{}^{b} \ , \quad 
\nabla_a \L^{\underline{b}} := e_a \L^{\underline{b}} + \omega_a{}^{\underline{c}} \xi^d f_{d\underline{c}}{}^{\underline{b}}
+ \omega_a{}^{\underline{c}} \L^{\underline{d}} f_{\underline{d} \underline{c}}{}^{\underline{b}} \ ,
\ee
where the structure constants are defined by
\be 
[X_{\underline{a}} , P_b ] = - f_{\underline{a} b}{}^{\underline{c}} X_{\underline{c}}
- f_{\underline{a} b}{}^{c} P_c \ , \quad 
[X_{\underline{a}} , X_{\underline{b}}] = - f_{\underline{a}\underline{b}}{}^{\underline{c}} X_{\underline{c}} \ .
\ee

The gauging procedure 
ensures that the generators 
$X_{\underline{a}}$ act on the 
covariant derivatives in the same way as they do on $P_a$, except with $P_a$ replaced by 
$\nabla_a$, while the covariant derivative algebra obeys commutation relations of the form
\be
[\nabla_a , \nabla_b] = -T_{ab}{}^c \nabla_c - \hf \RM_{ab}{}^{cd} M_{cd}
	- \RD_{ab} \mathbb D - \RK_{ab}{}^c K_c \ ,
\ee
where the curvatures and torsion are given by the form expressions:
\bsubeq
\begin{align}
T^a &= \hf e^c \wedge e^b \, T_{bc}{}^a = \rd e^a+ e^a \wedge b + e^b \wedge \omega_b{}^a \ , \\
\RM^{cd} &= \hf e^b \wedge e^a \RM_{ab}{}^{cd} = \rd \omega^{cd} + \omega^{ce} \wedge \omega_e{}^d - 4 e^{[c} \wedge \mathfrak{f}^{d]} \ , \\
\RD &= \hf e^b \wedge e^a \RD_{ab} = \rd b + 2 e^a \wedge \mathfrak{f}_a \ , \\
\RK^a &= \hf e^c \wedge e^b \RK_{bc}{}^a = \rd \mathfrak{f}^a - \mathfrak{f}^a \wedge b + \mathfrak{f}^b \wedge \omega_b{}^a \ .
\end{align}
\esubeq
The gravity gauge group acts on a tensor field $U$ (with indices suppressed) as 
\be \d_{\cK} U = \cK U \ .
\ee
We call a field $U$ satisfying $K_a U = 0$ and $\mathbb D U = \D U$ a primary 
field of dimension (or Weyl weight) $\D$.

To describe conformal gravity, one must impose some conformal constraints:
\be T_{ab}{}^c = 0 \ , \quad \eta^{bc} R(M)_{abcd} = 0 \ , \quad R(\bbD)_{ab} = 0 \ .
\ee
For $D > 3$, the Bianchi identities constrain the covariant derivative algebra to be of the form
\be [ \nabla_a , \nabla_b ] = 
- \hf C_{ab}{}^{cd} M_{cd} - \frac{1}{2 (D - 3)} \nabla^d C_{abcd} K^c \ ,
\ee
where $C_{abcd}$ is the Weyl tensor satisfying\footnote{
The symmetry property $C_{abcd} = C_{cdab}$ is not independent and follows from 
the others.}
\be C_{abcd} = C_{[ab][cd]} \ , \quad C_{[abc]d} = 0
\ee
and the Bianchi identity
\be \nabla_{[a} C_{bc]}{}^{de} = - \frac{2}{D - 3} \nabla_{f} C_{[ab}{}^{f [d} \d_{c]}^{e]}  \ .
\ee
The Weyl tensor $C_{ab}{}^{cd}$ proves to be a primary field.\footnote{This follows from 
considering $[K_a , [\nabla_b , \nabla_c]] = 2 [[K_a , \nabla_{[b} ] , \nabla_{c]}] = 0$.} This means 
that when the explicit expression for $\omega_a{}^{bc}$ is used dependence on $b_a$ drops out of 
the Weyl tensor.

One can always make use of the special conformal 
gauge freedom to choose a vanishing dilatation connection, $b_a = 0$. The covariant 
derivatives then take the form
\be \label{fextract}
\nabla_a = \cD_a - \frak{f}_a{}^b K_b \ , \quad \cD_a := e_a - \hf \omega_a{}^{bc} M_{bc} \ .
\ee 
In this gauge 
the Lorentz curvature
\be
\cR_{ab}{}^{cd} := 2 e_{[a}{}^m e_{b]}{}^n \partial_m \omega_{n}{}^{cd}
	- 2 \omega_{[a}{}^{cf} \omega_{b]f}{}^{d}
\ee
may be expressed as
\be  \cR_{ab}{}^{cd} = C_{ab}{}^{cd} - 8 \d_{[a}^{[c} \mathfrak{f}_{b]}{}^{d]}  \ .
\ee
One can then solve the special conformal connection in terms of the Lorentz curvature
\be \mathfrak{f}_{ab} = - \frac{1}{2 (D - 2)} \cR_{ab} + \frac{1}{4 (D - 1) (D - 2)} \eta_{ab} \cR \ , 
\ee
where we have defined
\be \cR_{ac} := \eta^{bd} \cR_{abcd} \ , \quad \cR := \eta^{ab} \cR_{ab} \ .
\ee
We will often refer to the procedure of setting $b_a = 0$ and introducing the covariant 
derivative $\cD_a$ as degauging.

It is worth mentioning that one can introduce new covariant derivatives by making use of a compensator $\phi$, 
which we choose to be primary and of dimension 2. 
One can construct the following covariant derivatives using the 
compensator
\be {\mathscr D}_a = \phi^{- \hf} \Big( \nabla_a 
+ \hf (\nabla^b \ln \phi) M_{ab}
- \hf (\nabla_a \ln \phi) \mathbb D
\Big) \ ,
\ee
which have the property that if $U$ is some conformally primary tensor 
field of some dimension then ${\mathscr D}_a U$ is as well. 
The covariant derivatives 
annihilate the compensator $\phi$, ${\mathscr D}_a \phi = 0$. 
When acting on primary fields they satisfy the algebra
\be [{\mathscr D}_a , {\mathscr D}_b] = - \hf {\mathscr R}_{ab}{}^{cd} M_{cd} \ ,
\ee
where
\be {\mathscr R}_{ab}{}^{cd} := \phi^{-1} C_{ab}{}^{cd}
+ \frac{4}{D - 2} \d_{[a}^{[c} {\mathscr R}_{b]}{}^{d]}
- \frac{2}{(D-1)(D-2)} \d_{[a}^{[c} \d_{b]}^{d]} {\mathscr R}
\ee
and
\begin{align}
{\mathscr R}_{ab} &:=
\hf \phi^{-1/2} (\nabla_{(a} \nabla_{b)} - \frac{1}{D} \eta_{ab} \Box) \phi^{- 1/2} 
+ \frac{D - 1}{D (D - 2)}\eta_{ab} \phi^{- (D + 2) /4} \Box \phi^{(D - 2)/4}
\ .
\end{align}
Here we have introduced the conformal d'Alembert operator $\Box := \nabla^a \nabla_a$. Upon 
degauging and 
imposing the gauge conditions $b_a = 0$ and $\phi =1$, one finds 
${\mathscr R}_{ab}{}^{cd}$ corresponds to the Lorentz curvature $\cR_{ab}{}^{cd}$.

In what follows we will specialize to the six dimensional case. We will find that all 
conformal gravity invariants can be constructed as
\be I = \int \rd^6 x \,e \, L \ , \quad K_a L = 0 \ , \quad \bbD L = 6 L \ ,  \label{confGravActForm}
\ee
where $L$ is a function of $C_{abcd}$, its covariant derivatives and possibly 
a compensator $\phi$ (but with $I$ possessing no dependence on $\phi$).


\subsection{The $C^3$ invariants}

Taking into account the symmetries of the Weyl tensor there are two inequivalent ways of 
contracting indices in the product of three Weyl tensors. These are as follows:
\be
L_{C^3}^{(1)} := C_{abcd} C^{aefd} C_{e}{}^{bc}{}_{f} \ , \quad L_{C^3}^{(2)} := C_{abcd} C^{cdef} C_{ef}{}^{ab} \ .
\ee
These lead to two inequivalent invariants 
$I^{({\rm i})}_{C^3} := \int \rd^6 x \,e \, L_{C^3}^{({\rm i})}$,  ${\rm i} = 1, 2 $.

It is worth noting that a special combination of the above invariants can be written in the following 
form:
\be  \label{C3speciaL}
- \frac{1}{8} \eps^{abcdef} \eps_{a'b'c'd'e'f'} C_{ab}{}^{a'b'} C_{cd}{}^{c'd'} C_{ef}{}^{e'f'}
 = 4 L_{C^3}^{(2)}
+ 8 L_{C^3}^{(1)} \ .
\ee
It will turn out that it is precisely this combination that permits a supersymmetric generalization.


\subsection{The $C \Box C$ invariant}

Considering the product of two Weyl tensors with 
two covariant derivatives, one finds the following primary
\bea
L_{C \Box C} := C^{abcd} \Box C_{abcd} 
+ \hf \nabla_e C_{abcd} \nabla^e C^{abcd}
+ \frac{8}{9} \nabla^d C_{abcd} \nabla_e C^{abce} \ ,
\eea
which leads to the corresponding invariant $I_{C \Box C} = \int \rd^6 x \,e \, L_{C \Box C}$.

Making use of the identity
\begin{align}
& C^{abcd} \Box C_{abcd} 
+ \hf \nabla_e C_{abcd} \nabla^e C^{abcd}
+ \frac{8}{9} \nabla^d C_{abcd} \nabla_e C^{abce} \non\\
&= \frac{1}{6} C^{abcd} \Box C_{abcd}
+ \hf \nabla_e \Big( C_{abcd} \nabla^e C^{abcd} 
+ \frac{16}{9} C^{abce} \nabla^d C_{abcd}
\Big) - \frac{4}{3} L_{C^3}^{(1)} + \frac{1}{3} L_{C^3}^{(2)}
\end{align}
and upon degauging (and removing a total derivative) one finds
\bea
I_{C \Box C} &=& \frac{1}{6} \int \rd^6 x \,e \, \Big[ C^{abce} \Big( \d_e^f \cD^2 - 4 \cR_e{}^f + \frac{6}{5} \d_e^f \cR \Big) C_{abcf} 
- 8 L_{C^3}^{(1)} + 2 L_{C^3}^{(2)} \Big] \ ,
\eea
where $\cD^2 := \cD^a \cD_a$.


\subsection{The Euler invariant} \label{Einva}

The Euler invariant may be constructed most easily in the gauge $b_a = 0$. In this gauge 
we define the Euler invariant as
\bea 
\label{bosonicEuler}
\cE_6 &:=& - \frac{1}{8} \eps^{abcdef} \eps_{a'b'c'd'e'f'} {\cR}_{ab}{}^{a'b'} {\cR}_{cd}{}^{c'd'} {\cR}_{ef}{}^{e'f'} \non\\
&=& 
4 L_{C^3}^{(2)}
+ 8 L_{C^3}^{(1)}
- 6 C^{abcd} C_{abce} \cR_d{}^e 
+ \frac{6}{5} C^{abcd} C_{abcd} \cR \non\\
&&
+ 3 C_{abcd} \cR^{bd} \cR^{ac}
+ \frac{3}{2} \cR_a{}^b \cR_b{}^c \cR_c{}^a 
- \frac{27}{20} \cR^{ab} \cR_{ab} \cR
+ \frac{21}{100} \cR^3 \ .
\eea
Although one can use the above expression, we will instead look for an alternative 
description for the Euler invariant that is manifestly primary.

To begin with one can show that the following field\footnote{
This can be compared with the result in \cite{Wunsch} 
for primary covariants in six dimensions.}
\be
E_6 := \Big( \Box^3 - \frac{8}{3} (\nabla^{b} \nabla^{d} C_{abcd}) \nabla^{a} \nabla^{c} \Big) \ln \phi 
\label{LOGconstruction}
\ee
is primary. 
Furthermore, the corresponding invariant
\be
I_{\rm Euler} := \int \rd^6 x \,e \, E_6
\ee
does not actually depend on the compensator. To see this we make 
a reparametrization
\be \phi \rightarrow e^{- \s} \phi \ , \quad \bbD \s = 0 \ , \label{reparametrisationPhis}
\ee
which induces the shift
\be E_6 \rightarrow E_6 
- \Big( \Box^3 - \frac{8}{3} (\nabla^{b} \nabla^{d} C_{abcd}) \nabla^{a} \nabla^{c} \Big) \s \ . \label{shiftSiGma}
\ee
At this point it is tempting to think that the term involving $\Box^3 \s$ is a total derivative. However, 
integration of $\nabla_a$ is complicated by the presence of the special conformal connection 
and it is usually easier to work in the gauge $b_a = 0$ to arrange a total derivative. We now proceed 
to do this and show that $E_6$ shifts by a total derivative under the reprarametrization \eqref{reparametrisationPhis}.

In the gauge $b_a = 0$ we find the following results:
\bsubeq
\bea
- \frac{8}{3} (\nabla^{b} \nabla^{d} C_{abcd}) \nabla^{a} \nabla^{c} \s &=& \frac{32}{3} \frak{f}^{ac} (\cD^b \s) \cD^d C_{abcd} + {\rm total \ derivative} \ , \\
\Box^3 \s &=& - \frac{32}{3} \frak{f}^{ac} (\cD^b \s) \cD^d C_{abcd} + {\rm total \ derivative} \ ,
\eea
\esubeq
where we made use of the identities
\be \cD_{[a} f_{b] c} = \hf R(K)_{abc} = \frac{1}{12} \nabla^d C_{abcd} \ , \quad \cD_a \frak{f}_b{}^{b} = \cD_{b} \frak{f}_a{}^b \ .
\ee
It is now straightforward to see that the shift in \eqref{shiftSiGma} is a total derivative 
and $I_{\rm Euler}$ is invariant under reparametrizations of $\phi$.

Since $I_{\rm Euler}$ does not depend on $\phi$, we are free to set 
$\phi = 1$, and since this condition breaks dilatation symmetry it is natural to work in the 
gauge $b_a = 0$. To do this consistently one must first extract the special conformal connection 
as in \eqref{fextract} before imposing the gauge conditions $\phi = 1$ and $b_a = 0$. Non-trivial terms survive 
which derive from where the dilatation generator acts on $\ln \phi$. One finds 
the following:
\begin{align}
- \frac{8}{3} (\nabla^{b} \nabla^{d} C_{abcd}) \nabla^{a} \nabla^{c} \ln \phi 
&=
- 2 L_{C \Box C}
- 2 C^{abce} C_{abcd} \cR^d{}_e 
+ \frac{2}{5} C^{abcd} C_{abcd} \cR \non\\
&
+ C_{abcd} \cR^{bd} \cR^{ac} 
- 4 L_{C^3}^{(1)}
+ L_{C^3}^{(2)} + {\rm total \ derivative} \ , \\
\Box^3 \ln \phi 
&= 
\frac{1}{2} \cR_a{}^b \cR_b{}^c \cR_c{}^a 
- \frac{9}{20} \cR^{ab} \cR_{ab} \cR 
+ \frac{7}{100} \cR^3 \non\\
&+ {\rm total \ derivative} \ .
\end{align}
Finally, it follows from the above that
\bea
E_6 &=&  
\frac{1}{3} \cE_6 - \frac{20}{3} L_{C^3}^{(1)} - \frac{1}{3} L_{C^3}^{(2)}  - 2 L_{C \Box C} + {\rm total \ derivatives} \ .
\eea
Interestingly, we find that besides the construction \eqref{LOGconstruction} containing the Euler invariant $\cE_6$, $E_6$ 
also involves the other conformal invariants.


\section{$\cN=(1,0)$ conformal superspace} 
\label{setup}

Conformal superspace in lower dimensions \cite{Butter4DN=1, Butter4DN=2, BKNT-M1, BKNT-M5D} 
possesses the following key properties: (i) it gauges the entire superconformal algebra; (ii) the
curvature and torsion tensors may be expressed in terms of a single primary superfield;
and (iii) the algebra obeys the same basic constraints as those of super Yang-Mills
theory. In this section, as in the lower dimensional cases, we will make use of these properties to develop the 
conformal superspace formulation for $\cN = (1, 0)$ conformal supergravity in six dimensions.
We will firstly give the superconformal algebra and describe the geometric setup for 
conformal superspace. We then constrain the geometry to 
describe conformal supergravity by constraining its covariant derivative algebra to be 
expressed in terms of a single primary superfield, the super-Weyl tensor.


\subsection{The superconformal algebra}

The 6D $\cN = (1, 0)$ superconformal algebra naturally originates as the algebra of 
Killing supervector fields of 6D $\cN = (1, 0)$ Minkowski superspace \cite{Park98}, 
see Appendix \ref{KVF} for the technical details. Below we simply summarize the 
(anti-)commutation relations of generators corresponding to the superconformal algebra.

The bosonic part of the 6D $\cN = (1, 0)$ superconformal algebra contains the translation ($P_{a}$), Lorentz ($M_{ab}$), 
special conformal ($K_{a}$), dilatation ($\mathbb{D}$) and $\rm SU(2)$ generators ($J_{ij}$), where 
$a, b = 0 , 1 , 2 , 3 , 4, 5$ and $i , j = \underline{1} , \underline{2}$. 
Their algebra is
\bsubeq
\begin{align} [M_{a b} , M_{c d}] &= 2 \eta_{c [a} M_{b] d} - 2 \eta_{d [ a} M_{b ] c} \ , \\
[M_{a b}, P_c] &= 2 \eta_{c [a} P_{b]} \ , \quad [\mathbb{D} , P_a] = P_a \ , \\
[M_{a b} , K_c] &= 2 \eta_{c [a} K_{b]} \ , \quad [\mathbb{D} , K_a] = - K_{a} \ , \\
[K_a , P_b] &= 2 \eta_{a b} {\mathbb D} + 2 M_{ab} \ , \\
[J^{ij} , J^{kl}] &= \eps^{k(i} J^{j) l} + \eps^{l (i} J^{j) k} \ ,
\end{align}
\esubeq
with all other commutators vanishing. The $\cN = (1, 0)$ superconformal algebra is obtained by extending the 
translation generator to $P_A = (P_a , Q_\a^i)$ and the special conformal generator to
$K^A = (K^a , S^\a_i)$.\footnote{
For our spinor conventions and notation we refer the reader to Appendix \ref{NC}.} 
The fermionic generator $Q_\a^i$ obeys the algebra
\bsubeq \label{SCA}
\begin{align} \{ Q_\a^i , Q_\b^j \} &= - 2 \ri \eps^{ij} (\g^{c})_{\a\b} P_{c} \ , \quad [Q_\a^i , P_a ] = 0 \ , \quad [{\mathbb D} , Q_\a^i ] = \hf Q_\a^i \ , \\
[M_{ab} , Q_\g^k ] &= - \hf (\g_{ab})_\g{}^\d Q_\d^k \ , \quad [J^{ij} , Q_\a^k ] = \eps^{k (i} Q_\a^{j)} \ ,
\end{align}
\esubeq
while the generator $S^\a_i$ obeys the algebra
\bsubeq\label{SCA2}
\begin{align} \{ S^\a_i , S^\b_j \} &= - 2 \ri \eps_{ij} (\tilde{\g}^{c})^{\a\b} K_{c} \ , \quad [S^\a_i , K_a ] = 0 \ , \quad [{\mathbb D} , S^\a_i ] = - \hf S^\a_i~, \\
[M_{ab} , S^\g_k ] &= \hf (\g_{ab})_\d{}^\g S^\d_k  \ , \quad [J^{ij} , S^\a_k ] = \d_k^{(i} S_\a^{j)} \ ,
\end{align}
\esubeq
Finally, the (anti-)commutators of $K^{A}$ with $P_A$ are
\bsubeq\label{SCA3}
\begin{align} [ K_a , Q_\a^i ] &= - \ri (\g_a)_{\a\b} S^{\b i} \ , \quad [S^\a_i , P_a ] = - \ri (\tilde{\g}_a)^{\a\b} Q_{\b i} \ , \\
\{ S^\a_i , Q_\b^j \} &= 2 \d^\a_\b \d^j_i \mathbb D - 4 \d^j_i M_\b{}^\a + 8 \d^\a_\b J_i{}^j \ ,
\end{align}
\esubeq
where we introduced $M_\a{}^\b = - \frac{1}{4} (\g^{ab})_\a{}^\b M_{ab}$. Note that $M_\a{}^\b$ acts on $Q_\g^k$ and 
$S^\g_k$ as follows
\be 
[ M_\a{}^\b , Q_\g^k ] = - \d_\g^\b Q_\a^k + \frac{1}{4} \d^\b_\a Q_\g^k \ , \quad 
[ M_\a{}^\b , S^\g_k ] = \d_\a^\g S^\b_k - \frac{1}{4} \d^\b_\a S^\g_k \ .
\ee


\subsection{Gauging the superconformal algebra} \label{Gauging}

To perform the gauging of the superconformal algebra we follow closely 
the approach given in \cite{Butter4DN=1, Butter4DN=2, BKNT-M1, BKNT-M5D}. Below 
we will give the salient details of the geometry.

We introduce a curved 6D $\cN = (1, 0)$ superspace
 $\cM^{6|8}$ parametrized by
local bosonic $(x)$ and fermionic coordinates $(\theta_i)$, $z^M = (x^{m}, \ \q^\mu_i)$, 
where $m = 0, 1, 2, 3, 4, 5$, $\mu = 1, \cdots,4$ and $i = \1, \2$. 
We associate with each generator $X_{\underline{a}} = (M_{ab} , J_{ij}, \mathbb D , S^\g_k , K^c)$ 
a connection one-form $\omega^{\underline{a}} = (\Omega^{ab} , \Phi^{ij}, B , \frak{F}_\g^k , \frak{F}_c)
= \rd z^{M} \omega_M{}^{\underline{a}}$ and with $P_A$ the vielbein $E^A = (E^\a_i , E^a)$. 
They are used to construct the covariant derivatives, which 
have the form
\be
\nabla_A  
= E_A - \hf \Omega_A{}^{ab} M_{ab} - \Phi_A{}^{kl} J_{kl} - B_A \mathbb D
	- \mathfrak{F}_A{}_B K^B \ .
\ee
Here $E_A = E_A{}^M \partial_M$ is the inverse vielbein. 
The action of the generators on the covariant derivatives 
resembles that for the $P_A$ generators given in \eqref{SCA}.

The supergravity gauge group is generated by local transformations of the form
\be
\delta_\cK \nabla_A = [\cK,\nabla_A] \ , \label{SUGRAtransmations}
\ee
where $\cK = \xi^C \nabla_C + \hf \L^{cd} M_{cd} + \L^{kl} J_{kl} + \s \mathbb D 
+ \L_A K^A$,
and the gauge parameters satisfy natural reality conditions. In applying
eq. \eqref{SUGRAtransmations}, one interprets the following
\be \nabla_A \xi^B := E_A \xi^B + \omega_A{}^{\underline{c}} \xi^D f_{D \underline{c}}{}^{B} \ , \quad 
\nabla_A \L^{\underline{b}} := E_A \L^{\underline{b}}
+ \omega_A{}^{\underline{c}} \xi^D f_{D \underline{c}}{}^{\underline{b}}
+ \omega_A{}^{\underline{c}} \L^{\underline{d}} f_{\underline{d}\underline{c}}{}^{\underline{b}} \ ,
\ee
where the structure constants are defined as
\be [X_{\underline{a}}, X_{\underline{b}} \}
= - f_{\underline{a} \underline{b}}{}^{\underline{c}} X_{\underline{c}} \ , \quad
[X_{\underline{a}} , \nabla_{B} \}
= - f_{\underline{a}B}{}^{C} \nabla_C - f_{\underline{a} B}{}^{\underline{c}} X_{\underline{c}} \ . 
\ee

The covariant derivatives satisfy the (anti-)commutation relations
\begin{align}
[ \nabla_A , \nabla_B \}
	&= -\scT_{AB}{}^C \nabla_C
	- \frac{1}{2} \sRM_{AB}{}^{cd} M_{cd}
	- \sRJ_{AB}{}^{kl} J_{kl}
	\non \\ & \quad
	- \sRD_{AB} \mathbb D
	- \sRS_{AB}{}_\g^k S^\g_k
	- \sRK_{AB}{}_c K^c~,
\end{align}
where the torsion and curvature tensors are given by
\begin{subequations} \label{torCurExp}
\bea
\scT^a &=& \rd E^a + E^b \wedge \Omega_b{}^a + E^a \wedge B \ , \\
\scT{}^\a_i &=& \rd E^\a_i 
+ E^\b_i \wedge \Omega_\b{}^\a 
+ \hf E^\a_i \wedge B - E^{\a j} \wedge \Phi_{ji} 
- \ri \, E^c \wedge \mathfrak{F}_{\b i} (\tilde{\g}_c)^{\a\b} \ ,~~~~~~~~~~~ \\
\sRD &=& \rd B + 2 E^a \wedge \mathfrak{F}_a + 2 E^\a_i \wedge \mathfrak{F}_\a^i \ , \\
\sRM^{ab} &=& \rd \Omega^{ab} + \Omega^{ac} \wedge \Omega_c{}^b 
- 4 E^{[a} \wedge \mathfrak{F}^{b]} 
+ 2 E^\a_j \wedge \mathfrak{F}_\b^j (\g^{ab})_\a{}^\b \ , \\
\sRJ^{ij} &=& \rd \Phi^{ij} - \Phi^{k (i} \wedge \Phi^{j)}{}_k - 8 E^{\a (i} \wedge \mathfrak{F}_{\a}^{j)} \ , \\
\sRK^a &=& \rd \mathfrak{F}^a 
+ \mathfrak{F}^b \wedge \Omega_b{}^a 
- \mathfrak{F}^a \wedge B 
- \ri \mathfrak{F}_\a^k \wedge \mathfrak{F}_{\b k} (\tilde{\g}^a)^{\a\b} \ , \\
\sRS_\a^i &=& \rd \mathfrak{F}_\a^i 
- \mathfrak{F}_\b^i \wedge \Omega_\a{}^\b
- \hf \mathfrak{F}_\a^i \wedge B 
- \mathfrak{F}_\a^j \wedge \Phi_j{}^i 
- \ri E^{\b i} \wedge \mathfrak{F}^c (\g_c)_{\a\b}    \ .
\eea
\end{subequations}
The covariant derivatives satisfy the Bianchi identities
\be 0 = [ \nabla_A , [ \nabla_B , \nabla_C \} \} +
	\text{(graded cyclic permutations)} \ .
\ee

A superfield $U$ is said to be \emph{primary} if it is annihilated by the special
conformal generators, $K^A U = 0$.
From the algebra \eqref{SCA2}, we see that if a superfield is annihilated by $S$-supersymmetry 
it is necessarily primary. The superfield $U$ is said to have dimension (or Weyl weight) $\D$ if 
$\mathbb D U = \D U$.


\subsection{Conformal supergravity}
\label{sec:CSG}

In the conformal superspace approach to supergravity 
in four \cite{Butter4DN=1, Butter4DN=2}, three \cite{BKNT-M1} 
and five dimensions \cite{BKNT-M5D}, 
the entire covariant derivative algebra may be expressed in terms of a single primary superfield: the super-Weyl tensor for
$D > 3$ 
and the super Cotton tensor for $D=3$.
In six dimensions we will look for a similar solution in terms of a 
single primary superfield,
the super-Weyl tensor \cite{LT-M12}.

In the lower dimensional cases 
the appropriate constraints to describe conformal supergravity 
were such that the covariant derivative algebra obeyed the 
same constraints as the super Yang-Mills theory.
Guided by the structure of 6D $\cN=(1, 0)$ super Yang-Mills theory \cite{Siegel78,Nilsson,Wess81,HST}, we constrain the 
covariant derivative algebra as
\bsubeq
\bea
\{ \nabla_\a^i , \nabla_\b^j \} &=& - 2 \ri \eps^{ij} (\g^a)_{\a\b} \nabla_a \ , \\ 
\left[ \nabla_a , \nabla_\a^i \right] &=& (\g_a)_{\a\b} \cW^{\b i} \ ,
\eea
\esubeq
where $\cW^{\a i}$ is some primary dimension $3/2$ operator taking values in the superconformal algebra. 
The Bianchi identities give the commutator
\bea [\nabla_a , \nabla_b] = - \frac{\ri}{8} (\g_{ab})_\a{}^\b \{ \nabla_\b^k , \cW^\a_k \}
\eea
and the additional constraints
\be \{ \nabla_\a^{(i} , \cW^{\b j)} \} = \frac{1}{4} \d^\b_\a \{ \nabla_\g^{(i} , \cW^{\g j)} \} \label{WalbeBI}
\ , \quad \{ \nabla_\g^k , \cW^\g_k \} = 0 \ .
\ee

We constrain the form of the operator $\cW^{\a i}$ to be
\be 
\cW^{\a i} = 
W^{\a\b} \nabla_\b^i
+ \hf \cW(M)^{\a i ab} M_{ab}
+ \cW(J)^{\a i jk} J_{jk}
+ \cW(\mathbb D)^{\a i} \mathbb D
+ \cW(K)^{\a i}{}_B K^B
 \ ,
\ee
where $W^{\a\b}$ is the super-Weyl tensor \cite{LT-M12} which is a symmetric primary superfield of 
dimension 1. One can show that the Bianchi identities \eqref{WalbeBI} are identically satisfied for
\bea
\cW^{\a i} &=& 
W^{\a\b} \nabla_\b^i
+ \nabla_\g^i W^{\a\b} M_\b{}^\g
- \frac{1}{4} \nabla_\g^i W^{\b\g} M_\b{}^\a
+ \frac{1}{2} \nabla_{\b j} W^{\a\b} J^{ij}
+ \frac{1}{8} \nabla_\b^i W^{\a\b} \mathbb D \non\\
&&
- \frac{1}{16} \nabla_\b^j \nabla_\g^i W^{\a \g} S^\b_j
+ \frac{\ri}{2} \nabla_{\b\g} W^{\g\a} S^{\b i} \non\\
&&- \frac{1}{12} (\g^{ab})_\b{}^\g \nabla_b \big( \nabla_\g^i W^{\b \a} 
- \hf \d_\g^\a \nabla_\d^i W^{\b\d} \big) K_a
\eea
provided $W^{\a\b}$ satisfies
\bsubeq
\bea 
\nabla_\a^{(i} \nabla_{\b}^{j)} W^{\g\d} &=& - \d^{(\g}_{[\a} \nabla_{\b]}^{(i} \nabla_{\r}^{j)} W^{\d) \r} \ , \\
\nabla_\a^k \nabla_{\g k} W^{\b\g} - \frac{1}{4} \d^\b_\a \nabla_\g^k \nabla_{\d k} W^{\g\d}
 &=& 8 \ri \nabla_{\a \g} W^{\g \b} \ .
\label{WBI}
\eea
\esubeq

It will be useful to introduce the dimension 3/2 superfields
\begin{gather}
X_\g^k{}^{\a\b} = -\frac{\ri}{4} \nabla_\g^k W^{\a\b} - \d^{(\a}_{\g} X^{\b) k} \ , \quad X^{\a i} := -\frac{\ri}{10} \nabla_{\b}^i W^{\a\b} \ , \label{dim3/2superfields}
\end{gather}
and the following higher dimension descendant superfields constructed 
from spinor derivatives of $W^{\a \b}$:
\bsubeq \label{YfielDs}
\begin{align}
Y_\a{}^\b{}^{ij} &:= - \frac{5}{2} \Big( \nabla_\a^{(i} X^{\b j)} - \frac{1}{4} \d^\b_\a \nabla_\g^{(i} X^{\g j)} \Big)
= - \frac{5}{2} \nabla_\a^{(i} X^{\b j)} \ , \\
Y &:= \frac{1}{4} \nabla_\g^k X^\g_k \ , \\
Y_{\a\b}{}^{\g\d} &:=
\nabla_{(\a}^k X_{\b) k}{}^{\g\d}
- \frac{1}{6} \d_\b^{(\g} \nabla_\r^k X_{\a k}{}^{\d) \r}
- \frac{1}{6} \d_\a^{(\g} \nabla_\r^k X_{\b k}{}^{\d) \r}
\ .
\end{align}
\esubeq
Note that $X_\g^k{}^{\a\b}$ is traceless, $Y_\a{}^{\b\, ij}$ is symmetric in its $\rm SU(2)$ indices and traceless in its
spinor indices, and $Y_{\a\b}{}^{\g\d}$ is separately symmetric in its
upper and lower spinor indices and traceless.

One can check that only the superfields \eqref{YfielDs} together with \eqref{dim3/2superfields} and 
their vector derivatives appear upon taking successive spinor derivatives of $W^{\a \b}$.
Specific relations we will need later are given below:
\bsubeq \label{eq:Wdervs}
\bea
\nabla_\a^i X^{\b j} &=& - \frac{2}{5} Y_\a{}^\b{}^{ij}
	- \frac{2}{5} \eps^{ij} \nabla_{\a \g} W^{\g\b}
- \hf \eps^{ij} \d_\a^\b Y \ , \\
\nabla_\a^i X_\b^j{}^{\g\d}
&=& 
\hf \d^{(\g}_\a Y_\b{}^{\d)}{}^{ij} - \frac{1}{10} \d_\b^{(\g} Y_\a{}^{\d)}{}^{ij}
- \hf \eps^{ij} Y_{\a\b}{}^{\g\d}
- \frac{1}{4} \eps^{ij} \nabla_{\a\b} W^{\g\d} \non\\
&&+ \frac{3}{20} \eps^{ij} \d_\b^{(\g} \nabla_{\a \r} W^{\d) \r}
- \frac{1}{4} \eps^{ij} \d^{(\g}_\a \nabla_{\b \r} W^{\d) \r} \ , \\
\nabla_\a^i Y &=& - 2 \ri \nabla_{\a \b} X^{\b i} \ , \\
\nabla_\g^k Y_{\a}{}^\b{}^{ij} &=& 
\frac{2}{3} \eps^{k (i} \Big( 
- 8 \ri \nabla_{\g\d} X_\a^{j)}{}^{\d \b}
- 4 \ri \nabla_{\a\d} X_\g^{j)}{}^{\d\b}
+ 3 \ri \nabla_{\g\a} X^{\b j)} \non\\
&&+ 3 \ri \d^\b_\g \nabla_{\a \d} X^{\d j)}
- \frac{3 \ri}{2} \d^\b_\a \nabla_{\g\d} X^{\d j)}
\Big) \ , \\
\nabla_\e^i Y_{\a\b}{}^{\g\d} &=& 
- 4 \ri \nabla_{\e (\a} X_{\b)}^l{}^{\g\d}
+ \frac{4 \ri}{3} \d^{(\g}_{(\a} \nabla_{\b) \r} X_\e^l{}^{\d ) \r} 
+ \frac{8 \ri}{3} \d^{(\g}_{(\a} \nabla_{|\e \r|} X_{\b)}^l{}^{\d) \r} \non\\
&&+ 8 \ri \d^{(\g}_{\e} \nabla_{\r (\a} X_{\b)}^l{}^{\d) \r} \ .
\eea
\esubeq
These equations guarantee that any number of spinor derivatives of $W^{\a\b}$
can always be rewritten in terms of $W^{\a\b}$, the superfields defined in
\eqref{dim3/2superfields} and \eqref{YfielDs},
and their vector derivatives.
The descendant superfields transform under $S$-supersymmetry as follows:
\bsubeq
\begin{align}
S^\a_i X_\b^j{}{}^{\g\d} &= -\ri \,\d^j_i \d^\a_\b W^{\g\d} 
+ \frac{2 \ri}{5} \d^j_i \d^{(\g}_{\b} W^{\d) \a} , \qquad
S^\a_i X^{\b j} = \frac{8\ri}{5} \d_i^j W^{\a\b} ~, \\
S^\g_k Y_\a{}^{\b}{}^{ij} &= \d^{(i}_k \Big( 
- 16 X_\a^{j)}{}^{\g\b} + 2 \d_\a^\b X^{\g j)}
- 8 \d^\g_\a X^{\b j)} \Big) \ , \\
S^\r_j Y_{\a\b}{}^{\g\d} &= 24 \Big( \d^\r_{(\a} X_{\b) j}{}^{\g\d}
- \frac{1}{3} \d^{(\g}_{(\a} X_{\b) j}{}^{\d) \r} \Big) \ , \quad S^\a_i Y = - 4 X^\a_i \ .
\end{align}
\esubeq

Expressing the covariant derivative algebra in terms of the descendant fields gives
\bsubeq
\bea
\{ \nabla_\a^i , \nabla_\b^j \} &=& - 2 \ri \eps^{ij} (\g^a)_{\a\b} \nabla_a \ , \\ 
\left[ \nabla_a , \nabla_\a^i \right] &=& (\g_a)_{\a\b} \Big(
W^{\b\g} \nabla_\g^i
+ 4 \ri\, X_\d^i{}^{\b\g} M_\g{}^\d
- \frac{\ri}{2} X^{\g i} M_\g{}^\b
- 5\ri\, X^\b_j J^{ij}
+ \frac{5\ri}{4} X^{\b i} \mathbb D \non\\
&&
+ \frac{\ri}{4} Y_\g{}^\b{}^{ij} S^\g_j
+ \frac{\ri}{4} \nabla_{\g\d} W^{\d\b} S^{\g i}
- \frac{5\ri}{16} Y S^{\b i} \non\\
&&
+ \frac{\ri}{3} (\g^{bc})_\d{}^\g \big( \nabla_b X_\g^i{}^{\d \b} - \frac{3}{4} \d^\b_\g \nabla_b X^{\d i} \big) K_c
\Big) \ .
\eea
\esubeq
An explicit expression for the remaining commutator
\bea
[\nabla_a , \nabla_b]
 &=& - \scT_{ab}{}^\g_k \nabla_\g^k 
 - \scT_{ab}{}^c \nabla_c 
 - \hf \sRM_{ab}{}^{cd} M_{cd}
 - \sRJ_{ab}{}^{kl} J_{kl}
 - \sRD_{ab} \bbD \non\\
 && - \sRS_{ab}{}_\g^k S^\g_k 
 - \sRK_{ab}{}_c K^c
\eea
follows from the Bianchi identities. For completeness, we 
provide the torsion and curvature components below:
\bsubeq
\bea
\scT_{ab}{}^c &=& - 4 W_{ab}{}^c \ , \\
\scT_{ab}{}^\gamma_k &=& (\gamma_{ab})_\beta{}^\alpha
	\Big(
	X_{\alpha k}{}^{\beta \gamma}
	- \frac{3}{4} \delta_\alpha{}^\gamma X^\beta_k
	\Big)~, \\
\sRM_{ab}{}^{cd} &=&
	Y_{ab}{}^{cd}
	- \delta_{[a}^c \delta_{b]}^{d} Y
	- 4 \nabla_{[a} W_{b]}{}^{cd}
	- 4   \nabla^{f} W_{f [b}{}^{[d} \delta_{a]}{}^{c]} \ ,\\
\sRJ_{ab}{}^{ij} &=& Y_{ab}{}^{ij} \ , \\
\sRD_{ab} &=& -2 \nabla^c W_{c ab}~, \\
\sRS_{a b}{}_{\gamma}^i &=&
	\ri (\gamma_{a b})_\beta{}^\alpha \Big(
	\tfrac{3}{16} \nabla_{\alpha\gamma}{X^{\beta i}}
	- \tfrac{3}{16} \delta_{\gamma}^{\beta} \nabla_{\alpha\delta}{X^{\delta i}}
	\non\\&&\qquad\qquad\qquad
	- \tfrac{1}{6} \nabla_{\alpha \delta}{X_{\gamma}^i{}^{\beta \delta}}
	- \tfrac{1}{3} \nabla_{\gamma \delta}{X_{\alpha}^i{}^{\beta \delta}}
	\Big) \ , \\
\sRK_{abc} &=& 
	\frac{1}{8} (\gamma_{c d [a})_{\alpha \beta} \nabla_{b]}{\nabla^{d}{W^{\alpha \beta}}}
	+ \frac{1}{4} \eta_{c [a} \nabla_{b]}{Y}
	+ \frac{1}{24} (\gamma_{a b})_\alpha{}^{\beta} (\gamma_{c d})_{\gamma}{}^{\delta} \nabla^{d}{Y_{\beta \delta}{}^{\alpha \gamma}}
	\eol &&
	+ \frac{1}{24} W^{\alpha \beta} Y_{\alpha \beta}{}^{\gamma \delta} (\gamma_{a b c})_{\gamma \delta}
	- \frac{1}{8} W^{\alpha \beta} Y_{\alpha \gamma}{}^{\delta \epsilon} (\gamma_c)_{\beta \delta} (\gamma_{a b})_{\epsilon}{}^{\gamma}
	\eol &&
	+ \frac{15\ri }{32} X^{\alpha k} X^{\beta}_k (\gamma_{a b c})_{\alpha \beta} 
	+ \frac{5\ri}{8} X^{\alpha k} X_{\alpha k}{}^{\beta \gamma} (\gamma_{a b c})_{\beta \gamma} 
	- \frac{5\ri}{4} X^{\alpha k} X_{\beta k}{}^{\gamma \delta} (\gamma_c)_{\alpha \gamma} (\gamma_{a b})_{\delta}{}^{\beta} 
	\eol &&
	+ \frac{\ri}{3} X_{\alpha}^k{}^{\beta \gamma} X_{\beta k}{}^{\alpha \delta} (\gamma_{a b c})_{\gamma \delta} 
	+ \ri X_{\alpha k}{}^{\beta \gamma} X_{\beta k}{}^{\delta \epsilon} (\gamma_c)_{\gamma \delta} (\gamma_{a b})_{\epsilon}{}^{\alpha}
	\eol &&
	+ \frac{5}{16} W^{\alpha \beta} (\gamma_c)_{\beta \delta} (\gamma_{[a})_{\alpha \gamma} \nabla_{b]}{W^{\gamma \delta}}
	- \frac{1}{32} W^{\alpha \beta} (\gamma_{c})_{\alpha \gamma} (\gamma_{a b d})_{\beta \delta} \nabla^{d}{W^{\gamma \delta}}
	\eol &&
	+ \frac{3}{32} W^{\alpha \beta} (\gamma_{a b c})_{\beta \delta} \nabla_{\alpha \gamma}{W^{\gamma \delta}}
	+ \frac{1}{16} W^{\alpha \beta} \eta_{c [a} (\gamma_{b]})_{\alpha \gamma} \nabla_{\beta \delta}{W^{\gamma \delta}} \ .
\eea
\esubeq

The component structure of 
the supergravity multiplet described by this superspace geometry 
can be identified with
the standard Weyl multiplet of $6D$ $\cN=(1,0)$ conformal supergravity \cite{BSVanP}.
The details of this will be presented in a future paper. Here we mainly
point out that the independent one-forms $e_m{}^a$, $\psi_m{}^{\alpha i}$, $b_m$,
and $V_m{}^{i j}$ in that approach coincide (up to conventions) with the $\q=0$
parts of the superspace one-forms $E_m{}^a$, $E_m{}^{\alpha i}$, $B_m$ and $\Phi_m{}^{i j}$,
respectively.
Similarly, the independent covariant fields $T_{abc}^-$, $\chi^{\alpha i}$, and
$D$ are given by the $\q=0$ parts of $W_{abc} = \frac{1}{8} (\g_{abc})_{\a\b} W^{\a\b}$, $X^{\alpha i}$, and $Y$.
The other components of the super-Weyl tensor $W^{\a\b}$ correspond to covariant
curvatures; for example, the $\q=0$ part of $Y_{ab}{}^{cd}$
is the traceless part of $R(M)_{ab}{}^{cd}$, which is
the supercovariant Weyl tensor.


\subsection{Introducing a compensator} \label{SU(2)viaCompensator}

An alternative formulation of conformal supergravity was given in \cite{LT-M12}, 
which we will refer to as SU(2) superpace. The formulation does not gauge the 
entire superconformal algebra and instead may be thought of as a 
gauge fixed version of the formulation introduced in the previous sections. Instead of 
applying the method of degauging used in \cite{Butter4DN=2, BKNT-M1, BKNT-M5D} 
we will make contact with SU(2) superspace by utilizing a compensator. Here we will 
develop the alternative approach advocated in lower dimensions in \cite{BN12, KNT-M15}, which makes clear 
how SU(2) superspace may be understood as conformal supergravity coupled to 
some compensator at the superspace level.

We introduce a primary superfield $X$ of dimension 2,
\be \mathbb D X = 2 X \ , \quad S^\a_i X = 0 \ .
\ee
The superfield can be 
used to furnish new spinor covariant derivatives, 
\bea
\mathscr D_\a^i &=& 
X^{- \frac{1}{4}} \Big( \nabla_\a^i
+ (\nabla_\b^i \ln X) M_\a{}^\b
- 2 (\nabla_\a^j \ln X) J_j{}^i
- \hf (\nabla_\a^i \ln X) \mathbb D \Big)
\ .
\eea
The covariant derivatives have been constructed to take a primary superfield
to another primary superfield of the same dimension. Note also that $X$ is annihilated 
by $\mathscr D_\a^i$, $\mathscr D_\a^i X = 0$.

When acting on a primary superfield, the algebra of
the covariant derivatives becomes\footnote{This agrees with 
the dimension 1 anticommutation relations of the covariant derivative 
algebra in \cite{LT-M12}.}
\bea
\{ \mathscr D_\a^i , \mathscr D_\b^j \}
&=& 
- 2  \ri \eps^{ij} {\mathscr D}_{\a\b}
- 4 \ri\eps^{ij} {\mathscr W}^{abc} (\g_a)_{\a\b} M_{bc} 
- 4 \ri\eps^{ij} {\mathscr N}^{abc} (\g_a)_{\a\b} M_{bc} 
+ 6 \ri \eps^{ij} \mathscr C_{\a\b}{}^{kl} J_{kl} \non\\
&& + 2 \ri {\mathscr C}_a{}^{ij} (\g^{abc})_{\a\b} M_{bc}
- 16 \ri {\mathscr N}_{\a\b} J^{ij}
\ ,
\eea
where
\bea
\mathscr D_{\a\b} &=& 
- \frac{\ri}{4} \{ \mathscr D_\a^k , \mathscr D_{\b k} \}
- 2 \mathscr N^{bcd} (\g_b)_{\a\b} M_{cd}
- 2 \mathscr W^{bcd} (\g_b)_{\a\b} M_{cd}
+ 3 \mathscr C_{\a\b}{}^{kl} J_{kl}
\eea
and we have introduced
\bsubeq
\bea
\mathscr C_{\a\b}{}^{ij} &:=& - \frac{\ri}{4} X^{- \frac{3}{2}} \nabla_\a^{(i} \nabla_\b^{j)} X
\ , \\
\mathscr N_{\a\b} &:=& - \frac{\ri}{16} X^{\frac{3}{2}} \nabla_{(\a}^{k} \nabla_{\b) k} X^{-2}
\ , \\
\mathscr W^{\a\b} &:=& X^{-\hf} W^{\a\b}
\ . 
\eea
\esubeq
Here we have introduced $\mathscr W^{\a\b}$ which is a rescaling of $W^{\a\b}$ so that 
it is inert under dilatations. The superfields $\mathscr C_{\a\b}{}^{ij}$ and 
$\mathscr N_{\a\b}$ are the only dimensionless primary combinations involving two spinor derivatives 
acting on $X$.
The super-Weyl transformations of \cite{LT-M12} correspond to a reparametrization of 
the compensator superfield, $X \rightarrow X \,e^{-2 \s}$.


\section{An action principle for the supersymmetric $C^3$ invariant} \label{actPrinc1}

Having developed conformal superspace in the previous section we are now in a position to address 
the problem of constructing conformal supergravity invariants. This will require an action 
principle capable of supporting such an invariant. In this section we expound such an action principle 
and show that it may be used to construct a supersymmetric $C^3$ invariant.

\subsection{Flat superspace actions and their generalization}
\label{sec:FlatActions}
Before discussing curved superspace actions, it is useful to briefly review action principles 
with manifest $\cN = (1,0)$ Poincar\'e supersymmetry. 
The simplest is the full superspace integral
\begin{align}\label{eq:FlatFS}
S = \int \rd^6x \, \rd^8\q\, \cL~,
\end{align}
where $\cL$ is an unconstrained real superfield. 
Because the Grassmann coordinates $\q^{\a i}$ are irreducible under the Lorentz and 
$R$-symmetry groups, there is no separate notion
of chiral superspace as in four dimensions. To construct smaller superspaces
involving a reduced set of $\q$'s, additional structure is needed.
The most well-known example is 6D $\cN=(1,0)$
harmonic superspace \cite{HSWest, Zupnik:1986da},
$\mathbb R^{6|8} \times S^2$,
where additional bosonic coordinates $u^{i\pm}$
are introduced to describe the coset space $S^2 = \rm SU(2) / U(1)$.\footnote{This
superspace is a natural extension of the 4D $\cN=2$ harmonic superspace 
\cite{GIKOS,GIOS}.}
Introducing a new basis for the 
Grassmann coordinates as $\q^{\a \pm} := u_i^\pm \q^{\a i}$,
one may construct an invariant action
\begin{align}\label{harmaction}
S = \int \rd^6x\, \rd u\, \rd^4\q^+\, 
\cL^{+4}
&= \int \rd^6x\, \rd u\, (D^-)^4 \cL^{+4}|_{\theta = 0}
~, \qquad
	D_\a^+ \cL^{+4} = 0~, 
\end{align}
where $D_\a^\pm := u_i^\pm D_\a^i$ and $\rd u$ is the invariant measure for $\rm SU(2)$. 
A special case is when
$\cL^{+4}$ is an $\cO(4)$ multiplet $C^{+4}$ with simple quartic dependence
on the harmonics, $C^{+4} \equiv u_i^+ u_j^+ u_k^+ u_l^+ C^{ijkl}$.
Its component action is given by\footnote{One can also have an action
principle with $C^{ijkl}$ obeying the weaker condition
$D_{(\a}^i D_{\b)}^j C^{klpq} = 0$. This leads to the action discussed
in eq. (4.72) of \cite{ALR14}.}
\begin{align}\label{eq:FlatO4action}
S &= \int \rd^6x\, \rd u\, (D^-)^4 C^{+4}|_{\theta = 0}
	= \frac{1}{5} \int \rd^6x\, (D^4)_{ijkl} C^{ijkl}|_{\theta = 0} ~, \eol
	(D^4)^{ijkl} &:= -\frac{1}{96} \veps^{\a\b\g\d} D_\a^{(i} D_\b^j D_\g^k D_\d^{l)}~,
	\qquad
	D_\a^{(i} C^{jklp)} = 0~.
\end{align}
For the similar case of 4D $\cN=2$ supersymmetry, the $\cO(4)$ multiplet
and associated action were introduced by \cite{SSW}.
It is clear that any full superspace action can be rewritten in this way
using $C^{+4} = (D^+)^4 \cL$. The converse is not always 
true within the family of local and gauge-invariant operators. More specifically, 
given an $\cO(4)$ multiplet $C^{+4}$, there always exists a harmonic-independent 
potential $\cL$ such that $C^{+4} = (D^+)^4 \cL$, 
as proved in Appendix G of \cite{BKNT-M5D} in the 5D $\cN=1$ case. 
However, such a potential $\cL$
 cannot always be defined as a local gauge-invariant operator. 
A simple example is when the $\cO(4)$ multiplet
is the product of two $\cO(2)$ multiplets.

Our task is to construct the conformal supergravity invariants,
so a natural step would be to generalize the above actions to curved superspace and
to choose the appropriate Lagrangians.
Both in SU(2) superspace \cite{LT-M12} and in conformal superspace, it is straightforward to generalize eq. \eqref{eq:FlatFS} to
\begin{align}\label{eq:CurvedFS}
S = \int \rd^6x\, \rd^8\q\, E\, \cL \ ,
\end{align}
where $E$ is the Berezinian (or superdeterminant) of the supervielbein. In order to
be invariant under the supergravity gauge transformations,
$\cL$ must be a conformal primary scalar superfield
of dimension two.
Unfortunately, there is no suitable Lagrangian that can be
built directly from the covariant fields of the Weyl multiplet. 
Furthermore, there is no obvious way to generalize \eqref{eq:FlatO4action} without
introducing a compensator field. The reason is $C^{+4}$ should clearly have
dimension four, but the analyticity condition $\nabla_\a^+ C^{+4} = 0$
cannot be conformally invariant, assuming $C^{+4}$ is a primary, unless $C^{+4}$
has dimension eight.
A natural step here would be to relax the assumption that $C^{+4}$ is itself primary
and instead consider it as a descendant of some other primary superfield.
One could imagine a number of ways of doing this.
In fact, we will discover an action principle in section \ref{actPrinc2} involving
such a non-primary $C^{+4}$.

For the moment, however, we will follow a different approach and attempt
to construct the actions as six-forms directly rather than as superspace
integrals.

\subsection{Primary closed six-forms in superspace}
\label{sec:PrimarySixForms}

While supersymmetric actions are frequently realized as integrals over the full superspace
or its invariant subspaces, there is an alternative construction involving 
the use of closed super $D$-forms 
\cite{Hasler, Ectoplasm, GGKS}.\footnote{The approach proves equivalent 
to the rheonomic formalism \cite{Castellani}.} 
For 6D $\cN = (1,0)$ superspace, we introduce a closed six-form $J$
\begin{align}
J = \frac{1}{6!}
\rd z^{M_6} \wedge \cdots \wedge \rd z^{M_1}  \,J_{M_1 \cdots M_6} 
\ ,\qquad
 \rd J = 0 ~.
\end{align}
(The closure condition is trivial on the spacetime $\cM^6$ since
there a six-form is a top form, but there are no top forms on the
supermanifold $\cM^{6|8}$ since $\rd \q^\mu_i$ commutes with itself.)
Such a closed superform leads immediately to the action principle
\bea 
S = \int_{\cM^6} i^* J = \int \rd^6 x \,e \,{}^* J |_{\q=0}\ , \qquad 
{}^*J := \frac{1}{6!} \eps^{mnpqrs} J_{mnpqrs} \ ,
\label{ectoS}
\eea
where $i:\cM^6 \rightarrow \cM^{6|8}$ is the inclusion map and
$i^*$ is its pullback, the effect of which is to project $\q^\mu_i =\rd \q^\mu_i=0$.
Closure of $J$ guarantees that the action is
invariant under 
general coordinate transformations of superspace.\footnote{Here 
we assume the general coordinate transformations are generated by a vector field 
$\x = \x^A E_A = \x^M \pa_M$ which vanishes at the boundary of $\cM^6$.}
In addition, the action must be invariant under all gauge transformations: for
conformal supergravity, this includes the standard superconformal transformations,
which form the subgroup $\cH$.
This implies that $J$ must transform into an exact form
\bea
 \d_{\cH} J =  \rd \Theta (\L^{\underline{a}} ) \ , \quad 
\L = \L^{\underline{a}} X_{\underline{a}}~.
\label{eq:deltaJExact}
\eea

A special case is when the closed six-form is itself invariant,
$\d_{\cH} J =  0$. This implies that if one instead decomposes $J$ in
the tangent frame,
\begin{align}
J &= \frac{1}{6 !} E^{A_6} \wedge 
\cdots
\wedge E^{A_1} 
J_{A_1 
\cdots
A_6}~,
\end{align}
the components $J_{A_1 \cdots A_6}$ transform covariantly and obey the
covariant constraints
\be \label{eq:CovClosure}
\nabla_{[A_1} J_{A_2 \cdots A_7 \} } + 3 T_{[A_1 A_2}{}^{B} J_{|B|A_3 \cdots A_7 \}} =  0 \ .
\ee
In particular, their $S$ and $K$ transformations are given by
\be
S^\b_j J_{a_1 \cdots a_{n}}{}_{\a_1}^{i_1}{\cdots}{}_{\a_{6 - n}}^{i_{6-n}} = 
- \ri \,n (\tilde{\g}_{[a_{1}})^{\b \g} J_{\g j}{}_{a_2 \cdots a_{n}]}{}_{\a_1}^{i_1}{\cdots}{}_{\a_{6- n}}^{i_{6-n}}{}  \ , \qquad
K^b J_{A_1 \cdots A_6} = 0~.\label{recurs}
\ee
Such superforms are called {\it primary}.

It follows from eq. \eqref{recurs} that the component of a primary superform 
with lowest dimension 
is a primary
superfield, so it is
natural to ask what primary constraints are compatible with the closure
conditions \eqref{eq:CovClosure}.
This general question was 
addressed by Arias et al. \cite{ALR14}
using 6D $\rm SU(2)$ superspace \cite{LT-M12}, and we will arrive at similar results
to theirs. First observe that the component of the superform $J$ with lowest dimension (which 
we will refer to as the lowest component of the superform) 
cannot be a scalar without
either that scalar being covariantly constant (which is forbidden by the superconformal algebra
due to its non-vanishing dimension)
or the superform being exact.\footnote{This is unlike what happens in four dimensions,
where one can construct the chiral action principle.}
This means we have to allow for the possibility that the lowest component 
carries some Lorentz and SU(2) indices. We let the lowest component of 
the superform be directly constructed in terms of the primary superfield
\be A_{\a_1 \cdots \a_n}{}^{\b_1 \cdots \b_m}{}^{k_1 \cdots k_p} = A_{(\a_1 \cdots \a_n)}{}^{\b_1 \cdots \b_m}{}^{(k_1 \cdots k_p)}
\ee
with dimension $\D$.
In analogy to the chiral action principle in 4D, we seek a primary constraint involving
one spinor derivative with totally symmetrized $\rm SU(2)$ indices,
$\nabla_\d^{(l} A_{\a_1 \cdots \a_n}{}^{\b_1 \cdots \b_m}{}^{k_1 \cdots k_p)}$.
Such constraints are natural: they appear in solving the first non-trivial Bianchi identity (if it is 
not identically satisfied) since the part symmetric in $\rm SU(2)$ indices cannot
be countered by the term proportional to the superspace torsion. We will suppose 
further that
\be \nabla_{(\a_1}^{(l} A_{\a_2 \cdots \a_{n+1})}{}^{\b_1 \cdots \b_m}{}^{k_1 \cdots k_p)} - {\rm traces} = 0\ , \label{genConstT}
\ee
where we subtract out all possible traces to render the result traceless in 
its spinor indices. Requiring the constraint to be primary implies 
\be 2 \D + 3 n + m - 4 p = 0 \ , \label{genNCondT}
\ee
which can only have solutions for $2 p \geq \D$.
Notice that the upper Lorentz indices are not assumed to be symmetric,
which generalizes some of the corresponding results of \cite{ALR14}. 
Remarkably, apart from the one degenerate case of the tensor multiplet,
all known closed primary superforms have underlying 
primary superfields satisfying a constraint of the form \eqref{genConstT} 
with the condition \eqref{genNCondT}.

We now seek to find a primary closed superform to act as an action principle supporting a supersymmetric 
$C^3$ invariant. 
Since we will want to set the superfield to be cubic in $W^{\a\b}$
and its spinor derivatives,
the underlying superfield should satisfy $\D \geq 3 + \frac{p}{2}$. 
Considering all the possible ways of embedding such a superfield into a (non-exact) closed six form 
leads one to consider a primary dimension $9/2$ superfield of the form $A_\a{}^{ijk}$ satisfying the constraint
\be \nabla_{(\a}^{(i} A_{\b)}{}^{jkl)} = 0 \ . \label{Aconst}
\ee
In fact, a superfield obeying this constraint was already used to construct a closed six-form 
in \cite{ALR14} in the context of 6D $\rm SU(2)$ superspace \cite{LT-M12};
such a superfield also appeared in the context of the anomalous current
multiplet \cite{HS98, KNS15}.
The resulting closed six-form is
\bsubeq \label{superformAction}
\bea 
J &=& \frac{1}{6 !} E^{A_6} \wedge \cdots\wedge E^{A_1} J_{A_1 \cdots A_6}~,\\
J_{abc}{}_\a^i{}_\b^j{}_\g^k &=& 
3 (\g_{abc})_{(\a\b} A_{\g)}{}^{ijk} 
\ , \\
J_{abcd}{}_\a^i{}_\b^j &=&
- \frac{\ri}{6} \eps_{abcdef} (\g^{ef})_{(\a}{}^{\g} S_{\b)\g}{}^{ij}
- \frac{\ri}{12} \eps_{abcdef} (\g^{efg})_{\a\b} (\tilde{\g}_g)^{\r\eta} E_{\r\eta}{}^{ij} 
\ ,\\
J_{abcde}{}_\a^i &=& \frac{\ri}{2} \eps_{abcdef} (\tilde{\g}^f)^{\b\g}
	(\Omega_{\b\g, \a}{}^i + \Omega_{\a \b , \g}{}^i)
\ , \\
J_{abcdef} &=& - \eps_{abcdef} F
\ ,
\eea
\esubeq
and all other components vanish. 
Here we have introduced the descendant superfields
\bsubeq
\bea S_{\a\b}{}^{ij} &:=& \frac{3}{4} \nabla_{(\a k} A_{\b)}{}^{ijk} \ , \quad
E_{\a\b}{}^{ij} := \frac{3}{4} \nabla_{[\a k} A_{\b]}{}^{ijk} \ , 
\\
\Omega_{\a\b, \g}{}^{i} &:=&
	\frac{\ri}{16} \nabla_{[\a j} \nabla_{\b] k} A_\g{}^{ijk} =
	\frac{\ri}{16} \nabla_{\a j} \nabla_{\b k} A_\g{}^{ijk} \ , \\
F &:=& \frac{1}{4!} \eps^{\a\b\g\d} \nabla_{\a i} \Omega_{\b\g ,}{}_\d^i 
= \frac{\ri}{2^4 4!}\eps^{\a\b\g\d} \nabla_{\a i} \nabla_{\b j} \nabla_{\g k} A_\d{}^{ijk} \
\ .
\label{4.15e}
\eea
\esubeq
Reality of the action implies that $\overline{A_\a{}^{ijk}} = A_{\a\, ijk}$,
and similarly for its descendants,
$\overline{E_{\a\b}{}^{ij}} = E_{\a\b\, ij}$,
$\overline{S_{\a\b}{}^{ij}} = S_{\a\b\, ij}$,
$\overline{\Omega_{\a\b, \g}{}^i} = \Omega_{\a\b, \g\, i}$,
and $\overline{F} = F$. These transform under $S$-supersymmetry as follows:
\begin{subequations}
\bea
S^\e_m S_{\a\b}{}^{ij} 
&=& - 24 \,\d^\e_{(\a} A_{\b)}{}^{ij}{}_m 
\ , \\
S^\e_m E_{\a\b}{}^{ij}
&=& - 18 \,\d^\e_{[\a} A_{\b]}{}^{ij}{}_m 
\ , \\
S^\d_l \Omega_{\a\b ,}{}_\g^k
&=& - 4 \ri\, \d^\d_{[\a} S_{\b]\g}{}^i{}_l
- 4 \ri \,\d^\d_{[\a} E_{\b]\g}{}^i{}_l
+ \frac{2 \ri}{3} \d^\d_\g E_{\a\b}{}^i{}_l \ , \\
S^\a_i F &=&
- 2 \,\eps^{\a\b\g\d} \Omega_{\b\g , \d i} \ .
\eea
\end{subequations}
Making use of these results one can check that the superform \eqref{superformAction} is 
primary.

It is worth mentioning that the closed six-form  \eqref{superformAction} 
may be derived by analogy with the construction of the closed four-form
\cite{GKT-M09} which describes 
the chiral action in 4D $\cN=2$ supergravity \cite{Muller89}.
Ref. \cite{GKT-M09} considered  the closed
four-form $ \o = F \wedge  F $, 
where $ F$ is the two-form field strength of an on-shell  U(1) vector multiplet. 
Under certain assumptions on the vector multiplet, it was shown that 
all components of $\o$ are expressed in terms of a single chiral $\cN=2$ superfield
$W^2$, with $W$ the chiral field strength of the vector multiplet. 
In the 6D $\cN=(1,0)$ case, 
one can consider the topological term $\Tr ( \bm F \wedge \bm F \wedge \bm F) $, where $\bm F$ is the two-form field strength of a YM 
multiplet, see Appendix \ref{YMmultiplet}. Rewriting 
the superform in terms of $A_\a{}^{ijk} \propto \eps_{\a\b\g\d} \Tr \big( \bm W^{\b (i} \bm W^{\g j} \bm W^{\d k)} \big)$, where $\bm W^{\a i}$ is 
the field strength of the Yang-Mills supermultiplet, and throwing away 
a covariantly exact piece one uncovers the structure of the superform $J$.
 

\subsection{The supersymmetric $C^3$ invariant}

In order to describe the supersymmetric $C^3$ invariant it is now necessary 
to construct a composite $A_\a{}^{ijk}$ out of the super-Weyl tensor. Since 
the invariant must contain 
a $C^3$ term 
and since the Weyl tensor 
is directly constructed out of the spacetime projection of the superfield $Y_{\a\b}{}^{\g\d}$, 
the composite $A_\a{}^{ijk}$ must be at least cubic in $W^{\a\b}$ and its descendants.
Taking into account the constraints on $A_\a{}^{ijk}$ gives 
the following unique solution:
\bea
A_\a{}^{ijk} 
&=& 5 \ri \eps_{\a\b\g\d} X^{\b(i} X^{\g j} X^{\d k)}
- 8 \ri \eps_{\a\b\g\d} X^{\b (i} X_{\a'}^j{}^{\g\b'} X_{\b'}^{k)}{}^{\d \a'}
+ \frac{64\ri }{3} \eps_{\a\b\g\d} X_{\a'}^{(i}{}^{\b\b'} X_{\b'}^j{}^{\g\g'} X_{\g'}^{k)}{}^{\d\a'} \non\\
&& + 4 \eps_{\a\b\g\d} Y_\r{}^{\b}{}^{(ij} X_\eta^{k)}{}^{\r \g} W^{\eta \d}
- 3 \eps_{\a\b\g\d} Y_\r{}^{\b}{}^{(ij} X^{\g k)} W^{\r \d} \ .
\label{eq:AW3}
\eea
In particular, one can check that the above superfield is primary and satisfies the constraint \eqref{Aconst}. 

The component reduction (although tedious) is straightforward and may be carried out 
similarly as in \cite{BN12}. Furthermore, one can readily verify that the 
action contains a $C^3$ term proportional to the combination \eqref{C3speciaL}. We leave the detailed analysis of the 
component action to a forthcoming paper.


\subsection{Other invariants} \label{actionsOtherThanA}

A natural question that one may ask is whether other invariants may be constructed using 
the same action principle. Specifically, can we construct another primary
composite $A_\a{}^{ijk}$ that is (for example) quadratic in the super-Weyl tensor $W^{\a\b}$?
Unfortunately, enumerating the possibilities it turns out that only the cubic
solution \eqref{eq:AW3} is possible. There are however certain composite primary
superfields that one can construct at dimension 3. These are
\bsubeq
\begin{align}
H^{\a\b \,ij} &=
W^{\g [\a} Y_{\g}{}^{\b]}{}^{ij}
+ 8\ri\, X_\g{}^{\d [\a (i} X_\d{}^{\b] \g j)}
-\frac{5\ri}{2} X^{[\a (i} X^{\b] j)}
\ ,
\label{eq:defH}
\\
H^{\a\b} &= Y W^{\a\b} + \frac{2}{7} Y_{\g\d}{}^{\a\b} W^{\g\d}
- \frac{\ri}{2} X^{\a k} X^{\b}_k
+ \frac{8\ri}{7} X_\g^k{}^{\d(\a} X_{\d k}{}^{\b) \g}
+ 4 \ri X^{\g k} X_{\g k}{}^{\a\b} \ .
\end{align}
\esubeq
It turns out that the first may be used to generate another action, 
which will be discussed in detail in the next section. Before moving on to the discussion there, it is 
worth illustrating the existence of the other action principle 
using the primary superform construction of this section.

The important property of 
eq. \eqref{eq:defH} (besides being primary) is that it satisfies the differential constraint\footnote{Notice that this constraint is 
a special case of eq. \eqref{genConstT}.}
\begin{align}\label{eq:BConstraint}
\nabla_\a^{(i} B^{\b\g \, j k)} = - 2 \ri \,\delta_\a^{[\b} \L^{\g]\, ijk}~,
\end{align}
with $B^{\a\b ij} = H^{\a\b ij}$ for some non-primary $\L^{\a ijk}$. One can check that it is not possible to construct a primary 
composite $A_\a{}^{ijk}$ directly from $B^{\a\b ij}$ 
with various covariant derivatives only. Despite this one can construct a composite $A_\a{}^{ijk}$ out of $B^{\a\b ij}$ with the 
use of a compensating 
supermultiplet. 
To demonstrate this we choose a compensating tensor multiplet $\Phi$, 
which satisfies the constraint
\be 
\nabla_\a^{(i} \nabla_\b^{j)} \Phi = 0 \ .
\label{4.21_tensor}
\ee
Then using the results of section \ref{SU(2)viaCompensator} (with $X = \Phi$), one can construct 
the following composite
\bea \label{acPrincbasedonB}
A_\a{}^{ijk}&=& 
 - \frac{1}{60} \Phi^{\frac{3}{4}} {\mathscr D}_{\a l} {\mathscr D}_\b^{(i} {\mathscr D}_\g^{j} B^{\b\g kl)}
+ 2 \ri \Phi^{\frac{3}{4}} {\mathscr N_{\a\b}} {\mathscr D}_\g^{(i} B^{\b\g jk)} 
- \frac{8 \ri}{3} \Phi^{\frac{3}{4}} ({\mathscr D}_\b^{(i} {\mathscr N_{\g\a}}) B^{\b\g jk)}
\non\\&&\qquad
+ \frac{2 \ri}{3} \Phi^{\frac{1}{4}} ({\mathscr D}_{\g}^{(i} W^{\b \g}) B_{\a\b}{}^{jk)}
+ a \Phi^{\frac{3}{4}} {\mathscr D}_\a^{(i} {\mathscr D}_{\b\g} B^{\b\g jk)} \ .
\eea
The last term involves a free parameter $a$ and generates an exact six-form, which 
may be removed. The composite $A_\a{}^{ijk}$ is primary and satisfies the differential 
constraint \eqref{Aconst}. As a result we can associate an action with any primary superfield satisfying eq. \eqref{eq:BConstraint}, 
and we therefore have an action principle based on $B^{\a\b ij}$.

The action principle based on $B^{\a\b ij}$, eq. \eqref{acPrincbasedonB}, can be used immediately to describe certain invariants. If we take 
$B^{\a\b ij} = H^{\a\b ij}$, the component action will contain a $C \Box C$ term. One can also 
construct a unique higher-derivative $F\Box F$ action for a non-abelian gauge theory by taking
\begin{align}
\label{actionFBoxF}
B^{\a \b\, i j} = \ri \Tr({\bm W}^{\alpha (i} {\bm W}^{\beta j)}) \ , 
\end{align}
where ${\bm W}^{\alpha i}$ is the field strength of the 6D $\cN=(1,0)$ Yang-Mills 
multiplet \cite{Siegel78,Nilsson,Wess81,HST}, see Appendix \ref{YMmultiplet} for details. The 
corresponding component action will contain a term of the form $\Tr({\bm F}_{ab} \Box {\bm F}^{ab})$ 
upon integrating by parts. 

It should be mentioned that in the rigid supersymmetric case 
the supersymmetric $F \Box F$ action was constructed in \cite{ISZ05}
within the harmonic superspace approach. Their result 
can also be recast as the $\cO(4)$ multiplet action 
\eqref{eq:FlatO4action} with 
\bea
C^{ijkl} \propto \Tr( {\bm X}^{(ij}  {\bm X}^{kl)})~,
\label{424}
\eea
where ${\bm X}^{ij} $ denotes the flat-superspace limit of the descendant
\eqref{C.8}. The interesting feature of the model proposed in \cite{ISZ05}
is that the operator $ {\bm X}^{ij} $ is not a primary superfield, 
but the action \eqref{eq:FlatO4action} based on \eqref{424} is 
superconformal.

It is important to point out that the action principle based on $B^{\a\b ij}$ may 
contain dependence on $\Phi$. Although we do not explicitly show this here, we expect that 
the action principle will be independent of the compensator. In the
the next section we show that such an action principle based on $B^{\a\b ij}$
exists without the need to introduce any compensator.

Before moving on we would like to mention one more application of the action principle based on a composite 
$A_\a{}^{ijk}$. Let $V^{\a i}$ be a prepotential for the tensor multiplet,\footnote{The prepotential 
for the tensor multiplet was introduced by Sokatchev in the framework of 
his harmonic-superspace formulation for 6D $\cN=(1,0) $
supergravity  \cite{Sokatchev:PhiPrepot}. More recently this prepotential has been
described in SU(2) superspace in \cite{LT-M12}.} 
\be \label{Tprepot}
\Phi = \nabla_{\a i} V^{\a i} \ , \quad \nabla_\a^{(i} V^{\b j)} = \frac{1}{4} \d_\a^\b \nabla_\d^{(i} V^{\d j)} \ , \quad K^A V^{\a i} = 0 \ .
\ee
It is defined modulo 
gauge transformations of the form
\be V^{\a i} \rightarrow V^{\a i} + W^{\a i} \ , 
\label{VgaugeTrans}
\ee
where $W^{\a i}$ is the field strength of an abelian vector multiplet, 
see Appendix \ref{YMmultiplet}. Using $V^{\a i} $
one can construct the following primary composite
\be A_\a{}^{ijk} = \eps_{\a\b\g\d} V^{\b (i} B^{\g\d jk)} \ . \label{VBactionPrinc}
\ee
It is simple to verify the differential constraint \eqref{Aconst}
by making use of \eqref{eq:BConstraint} and \eqref{Tprepot}. 
The action corresponding to the composite \eqref{VBactionPrinc} is 
invariant under arbitrary gauge  transformations \eqref{VgaugeTrans}
when $B^{\a\b ij}$ is further constrained as
\be [ \nabla_\a^{(i} , \nabla_{\b k} ] B^{\alpha \beta j) k} = - 8 \ri \nabla_{\a\b} B^{\a\b i j} \ ,
\label{addBConst}
\ee
which imposes a constraint on $B^{\a\b ij}$ to describe a closed 4-form \cite{ALR14}. 
Below we give two examples of gauge-invariant 
actions.

Our first example of a gauge-invariant action corresponds to the choice
 \eqref{actionFBoxF}.
In this case it is rather simple to see that a gauge transformation \eqref{VgaugeTrans} shifts the invariant by a 
topological term and the invariant contains the term $\Phi \Tr({\bm F}_{ab} {\bm F}^{ab})$.
Thus the action describes 
the non-Abelian vector multiplet coupled to the dilaton Weyl multiplet.
In the flat-superspace limit, 
the prepotential of the tensor compensator may be chosen as
$V^{\a i } \propto \q^{\a i}$. Then the top component \eqref{4.15e}
of the closed six-form 
 \eqref{superformAction} becomes
 \bea
 F \propto D_{\a i} D_{\b j} \Tr({\bm W}^{\alpha (i} {\bm W}^{\beta j)}) \ ,
 \eea
which is the  Lagrangian for the 6D $\cN=(1,0)$ super Yang-Mills theory 
postulated in \cite{HST}. Here we derived this Lagrangian from 
a more general action principle.

Our second example, derives from the fact that the constraint \eqref{addBConst} is 
satisfied for the composite \eqref{eq:defH}. In the case where $B^{\a\b ij} = H^{\a\b ij}$, eq. \eqref{VBactionPrinc}
may be seen to 
describe a 
supersymmetric Riemann curvature squared term  \cite{BSS1,BR}.


\section{An action principle for the 
supersymmetric $C \Box C$ invariant} \label{actPrinc2}

Although we have shown in the previous section that
one can construct a supersymmetric $C \Box C$ action 
with an explicit compensator field,
this has an obvious disadvantage. 
One would have to show that terms involving the compensator could be eliminated
by integrating by parts in order for it to be an invariant for minimal conformal supergravity.  
Due to the complexity involved in doing this, it would be better to have a compensator-independent approach, but as we have already
discussed, it seems impossible to generate an appropriate primary closed six-form.
This suggests that we should consider non-primary six-forms instead; however,
since these are rather more difficult to deal with, it would be 
helpful to know where to start looking.

Let us return to a point we raised earlier. The full superspace action \eqref{eq:CurvedFS}
is always a possible action principle, and it must correspond to some general
six-form action involving $\cL$ and its derivatives.
It turns out that its six-form \emph{cannot} be primary. The reason is that
if it were, then the lowest dimensional component would be
$S$-invariant and at least of dimension 3.
Now it is straightforward to investigate the $S$-transformation
properties of all higher components of $\cL$:
the only primary aside from $\cL$ itself appears at the $\q^2$ level,
\begin{align}\label{eq:BfromL}
B_{a}{}^{i j} = -\frac{\ri}{16} (\tilde\gamma_{a})^{\alpha\beta} \nabla_{\a}^{(i} \nabla_\b^{j)} \cL~.
\end{align}
(In particular, there is no primary at dimension 9/2 corresponding to $A_\a{}^{ijk}$
without introducing a compensator.)
We have denoted this descendant as $B_{a}{}^{i j}$ as it obeys the same constraint
\eqref{eq:BConstraint}
as the superfield $B^{\a\b\, i j} \equiv (\tilde\gamma^a)^{\a\b} B_a{}^{ij}$
introduced in the previous section.
Note however that it cannot be the bottom component of an
invariant six-form: it would have to be multiplied by six $E^\a_i$
to balance its dimension, but the Lorentz and SU(2) indices cannot be contracted appropriately.
This means that \emph{no corresponding primary six-form exists}.
Of course, it is not possible to construct an
invariant scalar $\cL$ from the superfields of the Weyl multiplet,
so what purpose does this observation serve? It turns out that one can build an
action principle upon a primary superfield $B_a{}^{ij}$ obeying certain properties
consistent with (but not implying) its derivation from a scalar superfield $\cL$.
In this way, $B_a{}^{ij}$ will lead to something analogous to the chiral action
principle of four dimensions.

The argument goes as follows.
Suppose we choose $\cL$ to be a tensor multiplet $\Phi$
subject to the constraint \eqref{4.21_tensor}.
Its superspace integral must vanish, 
\begin{align}
S = \int \rd^6x\, \rd^8\q\, E\, \F=0
\end{align}
since one can introduce the prepotential
$V^{\a i}$ for the tensor multiplet,  as in eq. \eqref{Tprepot},
and then integrate by parts.
Now the descendant $B_a{}^{ij}$ precisely vanishes
for a tensor multiplet, so it must be that that the six-form associated with
a general $\cL$ can be written purely in terms of the superfield
$B_a{}^{ij}$ and its derivatives.
This is analogous to the situation in four dimensions, where
a full $N\leq 2$ conformal superspace action can always be converted first
to a chiral superspace action using the chiral projection operator.
The converse is not true -- there are chiral Lagrangians that do not come from any
full superspace Lagrangian (at least not without introducing compensators).
Taking this analogy seriously, we conjecture that
any primary superfield $B_a{}^{ij}$ obeying the $S$-invariant constraint \eqref{eq:BConstraint}, 
which is consistent with \eqref{eq:BfromL}, must lead to an invariant action.

This proves to be precisely the action principle we need to describe the supersymmetric 
$C \Box C$ invariant. 
As a consequence of \eqref{eq:BConstraint}, one can show that 
\begin{align}\label{eq:dLdC}
\nabla_\a^{(i} \L^{\b j k l)} = \delta_\a{}^\b C^{ijkl}~, \qquad
\nabla_\a^{(i} C^{j k l p)} = 0~.
\end{align}
for non-primary superfields $\L^{\a\, ijk}$ and $C^{ijkl}$.
The superfield $C^{ijkl}$ is a non-primary version of
the $\cO(4)$ multiplet that we have already discussed in section \ref{sec:FlatActions},
and its $S$-transformation is exactly as needed to permit the second condition
of \eqref{eq:dLdC} to hold. This suggests that the six-form action principle should
begin with a term
\begin{align}\label{eq:BSixFormTop}
J = \frac{1}{6!}\, E^{a_1} \wedge \cdots \wedge E^{a_6} \,\veps_{a_1 \cdots a_6} \,F + \cdots~, \qquad
F = \frac{1}{5} (\nabla^4)_{i j k l} C^{ijkl}~,
\end{align}
providing a covariant version of the action principle \eqref{eq:FlatO4action}.
As already mentioned, we should not expect that the full six-form is primary.
Nevertheless, starting from the top component, one can iteratively
reconstruct the full six-form in a straightforward (albeit laborious) way.
The result turns out to include explicit $S$ and $K$ connections, which
makes $J$ transform into an exact form under those respective gauge
transformations.

We give the complete structure of this six-form in section \ref{sec:Baction}.
However, in order to better explain certain features of its construction, it helps
to describe the general properties of non-primary forms, especially if one
wishes to verify gauge invariance of the action. Section \ref{sec:NonPrimarySixForms}
is a self-contained discussion of this topic.


\subsection{Non-primary closed forms in superspace}
\label{sec:NonPrimarySixForms}
Let us begin with the following observation.
It has become apparent that superforms that are not invariant under
certain gauge symmetries nevertheless play an important role in constructing invariant
actions. These frequently involve Chern-Simons terms with bare connections:
recent examples have included the 4D and 5D linear multiplets \cite{BKN:LM, KN14:5DCS, BKNT-M5D},
3D $\cN \leq 6$ conformal supergravity \cite{BKNT-M2, KNT-M}, and
non-abelian $\cN\leq 4$ gauge theories \cite{KN:3DCS}.
However, such a geometric structure does not seem to be a necessary requirement.
For example, in the context of 4D $\cN=2$ conformal superspace,
bare $S$ and $K$ connections were recently observed
when constructing actions involving projective \cite{Butter:Proj}
and harmonic superfields \cite{Butter:Harm}. These were associated
with closed four-forms $J$ that transformed into exact forms under
$S$ and $K$ transformations. In this subsection, we will establish some
general properties of such non-primary closed forms in six dimensions.

Let $J$ be a closed super $p$-form. We assume it is invariant under Lorentz, Weyl, and
$\rm SU(2)$ transformations, but that it transforms under $K^A = (S^{\alpha}_i, K^a)$
transformations into an exact form. It is possible to expand $J$ in terms of
the vielbein $E^A$ and the $K$-connection $\mathfrak{F}_A$,
\begin{align}\label{eq:tempJExp}
J &= \frac{1}{p!} E^{A_1} \wedge \cdots \wedge E^{A_p} \,J_{A_p \cdots A_1}
	+ \frac{1}{(p-1)!} {\mathfrak F}_{A_1} \wedge E^{A_2} \wedge \cdots \wedge E^{A_{p}} J_{A_{p} \cdots A_2}{}^{A_1}
	\eol & \quad
	+ \cdots
	+ \frac{1}{p!} {\mathfrak F}_{A_1} \wedge \cdots \wedge {\mathfrak F}_{A_p} \,J^{A_p \cdots A_1}~,
\end{align}
so that the coefficient functions $J_{A_p \cdots A_{n+1}}{}^{A_n \cdots A_1}$ are
covariant superfields. Let us derive the conditions on these superfields so that
$\rd J = 0$.

Because $J$ is assumed to be invariant under Lorentz, Weyl, and $\rm SU(2)$
transformations, it is equivalent to analyze $\cD J = 0$ where
\begin{align}
\cD := \rd - \frac{1}{2} \Omega^{a b} M_{ab} - B \mathbb D - \Phi^{i j} J_{i j}
\end{align}
is covariant with respect to those symmetries. Using the definitions
\eqref{torCurExp} of the torsion tensor $T^A$ and $K$-curvature $R(K)_A$,
one verifies that
\begin{subequations}
\begin{align}
\cD E^A &= \frac{1}{2} E^B \wedge E^C T_{CB}{}^A + E^B \wedge {\mathfrak F}_C f^{C}{}_B{}^A~, \\
\cD {\mathfrak F}_A &= \frac{1}{2} E^B \wedge E^C R(K)_{CB A} + E^B \wedge {\mathfrak F}_C f^{C}{}_B{}_A
	+ \frac{1}{2} {\mathfrak F}_B \wedge {\mathfrak F}_C f^{C B}{}_A~,
\end{align}
\label{eq:tempdEF}
\end{subequations}
where the constants $f$ are the relevant structure constants appearing in the algebra
\begin{align}
[K^A, \nabla_B] &= -f^A{}_{B}{}^C \nabla_C - f^A{}_{B C} K^C + \text{other generators}~, \eol{}
[K^A, K^B] &= -f^{AB}{}_C K^C~.
\end{align}
From the definition of $\nabla_A$ one also has
\begin{align}\label{eq:tempdJcomp}
\cD J_{A_p \cdots A_{n+1}}{}^{A_{n} \cdots A_1} &=
	E^B \nabla_B J_{A_p \cdots A_{n+1}}{}^{A_{n} \cdots A_1}
	+ {\mathfrak F}_B K^B J_{A_p \cdots A_{n+1}}{}^{A_{n} \cdots A_1}~.
\end{align}

Now it is straightforward to analyze the conditions for closure on $J$.
These will be somewhat involved, so it is helpful to give a shorthand
approach that will allow us to compactly consider all equations at once.
We can introduce a generalized frame one-form $\cE^\cA = (E^A, {\mathfrak F}_A)$
and rewrite \eqref{eq:tempJExp} as\footnote{The notion of a
generalized frame appeared
naturally in the context of multiplets with central charge coupled to $\cN=2$
supergravity. There it facilitates the description of vector-tensor multiplets \cite{Novak:VT1, Novak:VT2}
and the construction of the linear multiplet action \cite{BKN:LM}.}  
\begin{align}
J &= \frac{1}{p!} \cE^{\cA_1} \wedge \cdots \wedge\cE^{\cA_p} \,J_{\cA_p \cdots \cA_1}~,
\end{align}
with the superfields $J_{\cA_p \cdots \cA_1}$ encapsulating those appearing in \eqref{eq:tempJExp}
in the obvious way.
This expansion formally treats the one-forms $E^A$ and ${\mathfrak F}_A$ on the same footing.
Imposing this democracy in the relations \eqref{eq:tempdEF} and \eqref{eq:tempdJcomp}
leads respectively to
\begin{align}
\cD \cE^\cA = \frac{1}{2} \cE^\cB \wedge \cE^\cC \cT_{\cC \cB}{}^{\cA}~, \qquad
\cD J_{A_p \cdots A_1} =
	\cE^\cB \nabla_\cB J_{\cA_p \cdots A_1} \ ,
\end{align}
where we have introduced $\nabla_{\cA} := (\nabla_A, K^A)$
and a tensor $\cT_{\cC \cB}{}^{\cA}$ defined as
\begin{alignat}{3}
\cT_{A B}{}^C &= T_{A B}{}^C~, &\quad 
\cT_{A}{}^{B C} &= f_{A}{}^{B C}~, &\quad
\cT^{AB}{}^C &= 0~, \eol
\cT_{A B}{}_C &= R(K)_{A B C}~, &\quad
\cT_{A}{}^{B}{}_C &= f_{A}{}^{B}{}_C~, &\quad
\cT^{AB}{}_C &= f^{AB}{}_C~.
\end{alignat}
Now it is immediately apparent that the
condition for closure on $J$ becomes
\begin{align}
\label{eq:GenClosure}
\nabla_{[\cA_{p+1}} J_{\cA_p \cdots \cA_1\}}
	+ \frac{p}{2} \cT_{[\cA_{p+1} \cA_p}{}^\cB J_{|\cB| \cA_{p-1} \cdots \cA_1\}} = 0 ~.
\end{align}

The above structure suggests the interpretation that we are enlarging the superspace
and introducing new coordinates associated with $K^A$ so that $\cE^\cA$ becomes
the new vielbein. From our perspective, this analogy is purely
a formal one -- we are \emph{not} introducing any new coordinates.
However, because the structure of the transformations is consistent with such a
possibility,\footnote{In more formal language, we could choose to work on the total
(super)space of the fiber bundle associated with $K$-transformations.}
many useful properties follow.
For example, the tensor $\cT$ can be interpreted as the generalized
torsion tensor of $\nabla_\cA$, that is
\begin{align}
[\nabla_\cA, \nabla_\cB] = -\cT_{\cA \cB}{}^{\cC} \nabla_{\cC} + \text{other generators}~.
\end{align}
Similarly, the $\delta_K$ transformations of the connections
$\cE^\cA = (E^A, {\mathfrak F}_A)$ and the covariant components $J_{\cA_p \cdots \cA_1}$
precisely satisfy a covariant form of Cartan's formula,
\begin{align}\label{eq:CartanForm}
\delta_K(\L) = \cD \imath_\L + \imath_\L \cD~,
\end{align}
where $\imath_\L$ is an antiderivation defined to act as
\begin{align}
\imath_\L {\mathfrak F}_A = \L_A~, \qquad
\imath_\L E^A = \imath_\L (\nabla_{\cA_{n}} \cdots \nabla_{\cA_{p+1}} J_{\cA_p \cdots \cA_1}) = 0~.
\end{align}
From these results, it is immediate to see that for a closed $p$-form \eqref{eq:tempJExp}
\begin{align}\label{eq:deltaKJ}
\delta_K(\L) J &= \cD \imath_\L J = \rd \imath_\L J~
\end{align}
which establishes that $J$ transforms as an exact form.

It is obvious that the class of primary superforms,
discussed in section \ref{sec:PrimarySixForms},
is simply one for which
no ${\mathfrak F}_A$ appears within the decomposition \eqref{eq:tempJExp}.
Then the closure condition \eqref{eq:GenClosure} amounts to two conditions:
\begin{subequations}
\begin{align}
\nabla_{[A_{p+1}} J_{A_p \cdots A_1\}}
	+ \frac{p}{2} \,T_{[A_{p+1} A_p}{}^B J_{|B| A_{p-1} \cdots A_1\}} = 0~, 
	\\
K^C J_{A_{p} \cdots A_1} + p \,f^C{}_{[ A_p}{}^B \,J_{|B| A_{p-1} \cdots A_1\}} = 0~.
\end{align}
\end{subequations}
The first is the usual covariant closure condition, and the second is the condition for $S$
and $K$-invariance (compare to eq. \eqref{recurs}). This illustrates how the single condition
\eqref{eq:GenClosure} concisely encodes both the conditions for
closure and for gauge invariance modulo an exact piece.


\subsection{A non-primary six-form action principle} \label{sec:Baction}
Now we turn to our specific goal of finding a non-primary six-form that begins
with the term \eqref{eq:BSixFormTop}.
Taking into account the closure conditions, one can deduce the structure of the remaining
terms. We use the definitions
\begin{subequations}
\begin{alignat}{3}
\L^{\a i j k} &:= \frac{\ri}{3} \nabla_{\b}{}^{(i} B^{\b\a j k)}~,&\quad
\L_{\a b}{}^i &:= \frac{2\ri}{3} \nabla_{\a j} B_b{}^{ij}~, \\
C^{ijkl} &:= \frac{1}{4} \nabla_\a^{(i} \L^{\a j k l)}~, &\,
C_\a{}^{\b i j} &:= \frac{3}{4} \nabla_{\a k} \L^{\b i j k}~, &\,
C_{a b} &:= \frac{1}{8} (\tilde\gamma_{a})^{\alpha \beta} \nabla_{\a k} \L_{\b b}{}^k~, \\
\rho_{\a}{}^{ijk} &:= -\frac{4\ri}{5} \nabla_{\a l} C^{i j k l}~, &\quad
\rho_{\alpha \beta}{}^{\gamma i} &:= -\frac{2\ri}{3} \nabla_{[\a j} C_{\beta]}{}^{\gamma i j}~,\\
E_{a}{}^{i j} &:= \frac{3}{16} (\tilde\gamma_a)^{\a\b} \nabla_{\alpha k} \rho_{\beta}{}^{ijk}~,
	\hspace{-1cm} \\
\Omega^{\alpha i} &:= \frac{\ri}{18} \nabla_{\beta j} E^{\b\a i j}~,&\,
F &:= \frac{1}{8} \nabla_{\a j} \Omega^{\a j} = \frac{1}{5} (\nabla^4)_{ijkl} C^{ijkl}~,
\hspace{-2cm}
\end{alignat}
\end{subequations}
with factors of $\ri$ chosen so that all fields obey
$\overline{\Psi^{i j \cdots}} = \Psi_{i j \cdots}$
where $\Psi$ carries any number of spinor indices.
In terms of these components, the action six-form may concisely be factorized as
\begin{align} \label{actPincB}
J = J_0 + {\mathfrak F}_{\a}^i \wedge J_S{}^{\a}_i + {\mathfrak F}_a \wedge J_K{}^a \ ,
\end{align}
where the six-form $J_0$ and the five-forms $J_S{}^{\a}_i$ and $J_K{}^a$ involve
only the supervielbein one-forms $E^A$.
The non-vanishing tangent-space components of $J_0$ are
\begin{align}
J_{0\,abc}{}_{\alpha}^i{}_\beta^j{}_\gamma^k &=
	- 3 (\gamma_{abc})_{(\a\b} \rho_{\g)}{}^{ijk}~, \eol
J_{0\, a b c d}{}_{\alpha}^i{}_\beta^j{} &=
	- \frac{8 \ri}{3} (\gamma_{[abc})_{\a\b} E_{d]}{}^{i j}~, \eol
J_{0\,a_1 a_2 a_3 a_4 a_5}{}_{\alpha}^i &= -\veps_{a_1 a_2 a_3 a_4 a_5 c}
\,  (\gamma^{c})_{\alpha \beta} \Big(
	\ri \Omega^{\beta i}
	- 8 \ri\, B_{a}{}^{i j} \nabla_{b}{X^{\gamma}_j} (\gamma^{a b})_{\gamma}{}^{\beta}
	\eol & \qquad\qquad
	+ \frac{32 \ri}{3} B_{a}{}^{i j} \,(\gamma^{a b})_{\gamma}{}^{\delta}\,
		\nabla_{b} X_{\delta j}{}^{\gamma \beta}
	- 3\ri\,  \Lambda^{\gamma i j k}\, Y_{\gamma}{}^{\beta}{}_{j k}
\Big)~, \eol
J_{0\,a_1 a_2 a_3 a_4 a_5 a_6} &= - \veps_{a_1 a_2 a_3 a_4 a_5 a_6} \,\Big(
F + 4\ri\, \Lambda_{\alpha b}{}^{k} (\gamma^{b c})_{\beta}{}^{\alpha} \nabla_{c}{X^{\beta}_k}
- \frac{16\ri}{3} \Lambda_{\alpha b}{}^{k} (\gamma^{b c})_{\beta}{}^{\gamma}
	\nabla_{c} X_{\gamma k}{}^{\beta \alpha}
\eol & \qquad \qquad
+ 2 B_{b\, i j} (\gamma^{b c})_{\alpha}{}^{\beta} \nabla_{c} Y_{\beta}{}^{\alpha i j}
- \frac{4}{3} C_{\beta}{}^{\alpha}{}_{i j} Y_{\alpha}{}^{\beta i j}  
\Big)~.
\end{align}
Note that there are some similarities between components of $J_0$ and those of
the $A_\a{}^{ijk}$ six-form \eqref{superformAction}. In
particular, the lowest dimensional component $\rho_\a{}^{ijk}$ of $J_0$
obeys the same differential constraint \eqref{Aconst} as $A_\a{}^{ijk}$; the difference
is that $\rho_\a{}^{ijk}$ is not primary but transforms into
$C^{ijkl}$ under $S$-supersymmetry.
The non-vanishing components of the five-forms $J_S{}^\alpha_i$ and $J_K{}^a$ are
simpler in structure and given by
\begin{subequations}
\begin{align}
J_S{}_{abc}{}_{\beta}^j{}_{\gamma}^k{}^{\alpha}_i &=
	24 \ri \,(\gamma_{abc})_{\b\g} \L^{\a\,j k}{}_i~, \\
J_S{}_{abcd}{}_{\beta}^j{}^{\alpha}_i &=
	\frac{8}{3} \veps_{abcdef} (\g^{ef})_\b{}^\g \, C_\g{}^{\a}{}^j{}_i~, \\
J_S{}_{abcde}{}^{\alpha}_i &= 
	\veps_{abcdef} (\tilde\gamma^f)^{\b\g} \rho_{\b\g}{}^{\a}_i~,
\end{align}
\end{subequations}
and
\begin{subequations}
\begin{align}
J_K{}_{bcd}{}_\a^i{}_\b^j{}^a &=
	-64 \ri (\g_{bcd})_{\a\b}\, B^{a\, i j}~,\\
J_K{}_{bcde}{}_\alpha^i{}^a &=
	8 \ri \,\veps_{bcdefg}\, (\g^{fg})_\a{}^\b\, \L_\b{}^{a \, i}~, \\
J_K{}_{bcdef}{}^a &= \veps_{bcdefg} (\tilde\g^g)^{\g\d} C_{\g\d}{}^{\a\b}\, (\g^a)_{\a\b}~.
\end{align}
\end{subequations}
They are essentially determined by the requirement that the full six-form
$J$ should transform as
\begin{align}
\delta_S J = - \rd (\Lambda_{S}{}_\alpha^i J_S{}^{\a}_i)~, \qquad
\delta_K J = - \rd (\Lambda_{K a} J_K{}^{a})~, \qquad
\end{align}
under $S$ and $K$ transformations, consistent with \eqref{eq:deltaKJ}.
Note that since $J$ is not primary, we may freely add
any exact form we choose to it. In particular, some of the terms in $J_S$ and $J_K$
can be removed by choosing such a form appropriately;
however, since it does not seem possible to eliminate
either $J_S$ or $J_K$ completely, we have not tried to simplify $J$ any
further.

Using this non-primary six-form, we can immediately construct the invariants 
corresponding respectively to the supersymmetric $C \Box C$ invariant and 
the supersymmetric $F \Box F$ actions. The first, as already
mentioned, involves choosing $B^{\a\b\, ij} = H^{\a\b\, i j}$ in \eqref{eq:defH}.
The leading components of the action can be deduced by observing that the non-primary
descendant $\cO(4)$ superfield is simply
\begin{align}
C^{ijkl} &= - \hf Y_\a{}^{\b (i j} Y_\b{}^{\a \,k l)}
\end{align}
from which the leading contributions to $F = \frac{1}{5} (\nabla^4)_{ijkl} C^{ijkl}$
may be determined. The term associated with the Weyl tensor is straightforward
to derive:
\begin{align}
F &=
	\frac{2}{9} (\nabla^d Y_{abcd})^2
	+ \cdots
	= \frac{2}{9} (\nabla^d R(M)_{abcd})^2 + \cdots~.
\end{align}
Note that even this leading term is not $K$-invariant, as one must include
the explicit $K$-connection terms in the six-form.
Removing a total derivative and higher order terms in the Weyl tensor leads to
\begin{align}
F &= - \frac{1}{12} R(M)_{abcd} \Box R(M)^{abcd} 
	+ \cdots~.
\end{align}
The second case, the supersymmetric $F \Box F$ action, involves the composite \eqref{actionFBoxF}.
Here one finds the non-primary $\cO(4)$ descendant superfield is
$C^{ijkl} = \Tr ({\bm X}^{(ij} {\bm X}^{kl)})$. As we have already noted, this is
precisely the harmonic superspace Lagrangian used in \cite{ISZ05} to construct this
invariant in flat space. At leading order, one finds the top component of the multiplet
is
\begin{align}
F = 2 \Tr(\nabla^{b} {\bm F}_{ba} \nabla_c {\bm F}^{ca}) + \cdots
	= -\Tr({\bm F}_{ab} \Box {\bm F}^{ab}) + \cdots \ ,
\end{align}
where we have discarded a total derivative and higher order terms.

The details of the component action corresponding to the supersymmetric $C \Box C$ and 
$F \Box F$ invariants will appear in a forthcoming paper.


\section{Discussion} \label{conclusion}

In this paper we have constructed two invariants for minimal conformal supergravity 
in six dimensions. 
These include 
the supersymmetric $C^3$ invariant described by the composite \eqref{eq:AW3} together with the 
action principle \eqref{superformAction}, as well as the supersymmeric 
$C \Box C$ invariant described by the composite \eqref{eq:defH} 
together with the action principle \eqref{actPincB}. 
The number of invariants constructed is consistent with the expectation that there should only be 
two in the case of $\cN=(1,0)$ local supersymmetry, see 
{\it e.g.} \cite{BT15}. 
However, it would be good to confirm 
that there does not remain another invariant. 
A rather simple way to answer
this question is to consider possible supercurrents of the Weyl multiplet.

In supersymmetric field theory,  the supercurrent is a supermultiplet containing
the energy-momentum tensor and the supersymmetry current(s), along with
some additional components such as the $R$-symmetry current. 
In the case of 6D $\cN = (1, 0)$ superconformal field theory, 
the supercurrent was described in \cite{HST} in Minkowski superspace. 
Its generalization to the curved case is  
described by a scalar primary superfield $\cJ$ of dimension 4 satisfying the 
differential constraint\footnote{This is the only possible curved extension 
of the flat case description in \cite{HST} provided $\cJ$ is primary.}
\be
\nabla_{[\alpha}^{(i} \nabla_{\beta}^{j} \nabla_{\gamma]}^{k)} \cJ = 0 \ . \label{Jconst}
\ee
When the superconformal theory is coupled to conformal supergravity,
the lowest component of $\cJ$ matches the variational derivative of the
action with respect to the highest dimension independent field
of the Weyl multiplet, which is the scalar auxiliary field $D$
as mentioned in section \ref{sec:CSG}.

We may now ask the following question: how many possible supercurrents can be built
purely from the super-Weyl tensor and its covariant derivatives?
The most general possible ansatz is
\bea
\cJ &=& c_1 \nabla^a \nabla_a Y
	+ c_2 Y^2
	+ \ri \,c_3 \,X^{\alpha i} \nabla_{\alpha \beta} X^\beta_i
	+ \ri \,c_4 \,X_\alpha^i{}^{\beta \gamma} \nabla_{\gamma \delta} X_{\beta i}{}^{\alpha \delta}
	+ c_5 \,Y_\alpha{}^\beta{}^{ij} Y_{\beta}{}^{\alpha}{}_{ij}
	\eol &&
	+ c_6 \,Y_{\alpha \beta}{}^{\gamma \delta} Y_{\gamma \delta}{}^{\alpha \beta}
	+ c_7 \,W^{\a\g} \nabla_{\a\b} \nabla_{\g\d} W^{\d\b}
	+ c_8 \nabla_{\b\a}W^{\a\g}  \nabla_{\g\d} W^{\d\b} 
	\eol &&
	+ c_9 \eps_{\a_1 \cdots \a_4} \eps_{\b_1 \cdots \b_4} W^{\a_1 \b_1} \cdots W^{\a_4 \b_4} \ ,
\eea
where $c_n$, $n = 1, \cdots 9$, are real coefficients. Requiring that $\cJ$ be primary
and satisfy the constraint \eqref{Jconst} yields a two-parameter family of possibilities,
\begin{alignat}{3}
c_3 &= -\frac{8}{3} c_2 - 5 c_1 ~, &\qquad
c_4 &= - \frac{32}{15} c_2 - 16 c_1 ~, &\quad
c_5 &= \frac{2}{15} c_2 + \frac{6}{5} c_1 ~, \eol
c_6 &=  \frac{2}{45} c_2 + \frac{1}{3} c_1~, &\quad
c_7 &= -\frac{2}{15} c_2 - \frac{1}{5} c_1~, &\quad
c_8 &= \frac{1}{2} c_7 = -\frac{1}{15} c_2 - \frac{1}{10} c_1~,
\eol
c_9 &= 0 \ ,
\end{alignat}
given here in terms of the coefficients $c_1$ and $c_2$. The family with $c_1 = 0$
corresponds to a supercurrent built from the cubic Weyl invariant,
whereas a combination with nonzero $c_1$ must correspond to
the quadratic Weyl invariant. There are no other possibilities,
so the two invariants we have constructed are the only ones. 

In section \ref{Einva} we discussed the Euler invariant, 
 eq. \eqref{bosonicEuler}. Here we briefly comment on
its extension to 
 the supersymmetric case. 
 It can naturally be introduced 
by first using the special conformal (and $S$-supersymmetry) transformations to 
gauge away
the dilatation connection entirely, $B_A =0$. 
It is now natural to perform the degauging procedure as in 
\cite{Butter4DN=1, Butter4DN=2, BKNT-M1, BKNT-M5D},  
and extract the special conformal connection $\frak{F}_A$ by introducing the {\it degauged} 
covariant derivatives $\cD_A := \nabla_A + \frak{F}_{AB} K^B$, 
with 
$\rm SO(5,1) \times SU(2)$ being the corresponding structure group.  
They satisfy 
(anti-)commutation relations of the form\footnote{The reader should be made aware that our 
notation for the curvature tensor coincides with the bosonic case, but they are not to be confused.}
\be [\cD_A , \cD_B \} = - \cT_{AB}{}^{C} \cD_{C} - \hf \cR_{AB}{}^{cd} M_{cd}
- \cR_{AB}{}^{kl} J_{kl} \ ,
\ee
where $\cT_{AB}{}^{C}$ is the torsion, and $\cR_{AB}{}^{cd}$ and $\cR_{AB}{}^{kl}$ are 
the Lorentz and SU(2) curvatures, respectively. 
A detailed analysis of the torsion and curvature tensors will be given elsewhere. 
The Euler  invariant is defined to be the closed six-form
\be 
\cE_6 = \frac{1}{8} \cR^{ab} \wedge \cR^{cd} \wedge \cR^{ef} \eps_{abcdef} \ ,
\qquad \rd \cE_6 =0~,
\ee
where $\cR^{cd} = \hf E^B \wedge E^A \cR_{AB}{}^{cd}$.

It may be seen that $\cE_6$ contains the same $C^3$ combination 
\eqref{C3speciaL} (modulo an overall coefficient)
  which originates in the closed six-form $J_{C^3}$ 
 describing the supersymmetric $C^3$ invariant, eq. \eqref{superformAction}. 
As a result, the closed six-form
\be \cE_6 + 12 J_{C^3} \ , \label{exp:SUSYC^3new}
\ee
does not contain any term involving only the Weyl tensor. All bosonic structures in the above invariant involve the Ricci tensor. However, it 
is not actually an independent invariant since we have only added a total derivative.

It was shown in section \ref{confGravity} that there exists a primary construction 
in terms of the logarithm of a 
compensator. Upon degauging the compensator it contains a linear combination of the conformal invariants. 
Although outside of the scope of this work it would be interesting to construct its supersymmetric 
extension. 

A detailed analysis of the component structure of the supergravity multiplet, as well as of the invariants for 6D $\cN=(1,0)$ conformal supergravity 
constructed, will be given in a forthcoming publication \cite{BNT-M}.
\\


\noindent
{\bf Acknowledgements:}\\
 We are grateful to Gabriele Tartaglino-Mazzucchelli for discussions and collaboration 
 at the early stage of this work.
The work of DB is supported in part by the ERC Advanced Grant no. 246974,
{\it ``Supersymmetry: a window to non-perturbative physics''}
and by the European Commission Marie Curie International Incoming Fellowship 
grant no. PIIF-GA-2012-627976. DB also thanks the organizers of the
2015 Simons Summer Workshop where the initial stages of this project were undertaken.
The work of SMK and ST is supported in part by the
Australian Research Council, project DP140103925. 
SMK and ST acknowledge hospitality of the
Arnold Sommerfeld Center for Theoretical Physics at 
Ludwig-Maximilians-Universit\"at in July 2015, 
where the program of constructing invariants for 6D $\cN=(1,0)$
conformal supergravity was initiated. ST and JN acknowledge support 
from GIF -- the German-Israeli Foundation for Scientific Research and Development. 
S.T. would like to thank A. Kashani-Poor and R. Minasian for hospitality at IHP in 
Paris and for useful discussions. 
JN would also like to acknowledge the hospitality of the School of Physics 
at The University of Western Australia where some of this work took place. SMK is grateful to 
Arkady Tseytlin for hospitality at Imperial College and for pointing out Ref. \cite{BT15}.


\appendix


\section{Notation and conventions} \label{NC}

We follow similar 6D notations and conventions as \cite{LT-M12},
with a few minor modifications. All relevant details are summarized here.

The Lorentzian metric is
$\eta_{ab} = \textrm{diag}(-1,1,1,1,1,1)$, the Levi-Civita tensor
$\veps_{abcdef}$ obeys $\veps_{012345} = -\veps^{012345} = 1$,
and the Levi-Civita tensor with world indices 
is given by $\eps^{mnpqrs} := \eps^{abcdef} e_a{}^m e_b{}^n e_c{}^p e_d{}^q e_e{}^r e_f{}^s$.

We exclusively use four component spinors in the body of the paper,
but it is useful to link these to eight component spinor conventions.
Our $8 \times 8$ Dirac matrices $\Gamma^a$ and charge conjugation matrix
$C$ obey
\begin{gather}
\{\G_a, \G_b\} = -2 \eta_{a b} \bm{1}~, \quad
(\G^a)^\dag = -\G_a~, \quad
C \G_a C^{-1} = -\G_a^T~, \eol
C^\dag C = \bm{1}~, \qquad C = C^T = C^*~.
\end{gather}
In particular, $\Gamma_a C^{-1}$ is antisymmetric. The chirality matrix
$\Gamma_*$ is defined by
\begin{align}
\Gamma_{[a} \Gamma_b \Gamma_c \Gamma_d \Gamma_e \Gamma_{f]} = \veps_{abcdef} \Gamma_*~.
\end{align}
As a consequence of the above conditions, one can show that
\begin{align}\label{eq:GammaReality}
\Gamma^a = B (\Gamma^a)^* B^{-1}~, \qquad B = \Gamma_* \Gamma_0 C^{-1}~.
\end{align}
The charge conjugate $\Psi^c$ of a Dirac spinor is conventionally defined by
\begin{align}
\bar\Psi \equiv \Psi^\dag \Gamma_0 =: (\Psi^c)^T C \qquad \implies \quad
\Psi^c = - \Gamma_0 C^{-1} \Psi^* = -\Gamma_* B \Psi^*~.
\end{align}
Because $B^* B = -1$,
charge conjugation is an involution only for objects with an even number of spinor indices,
so it is not possible to have Majorana spinors in six dimensions.
One can instead have a symplectic Majorana condition when the spinors possess an $\rm SU(2)$ index.
Conventionally this is denoted
\begin{align}\label{eq:SympMaj}
(\Psi_i)^c = \Psi^i \quad \implies \quad
\Psi^i = -\Gamma_0 C^{-1} (\Psi_i)^* = -\Gamma_* B (\Psi_i)^*
\end{align}
for a spinor of either chirality. We raise and lower $\rm SU(2)$ indices $i=\1,\2$
using the conventions
\begin{align}
\Psi^i = \eps^{i j} \Psi_j~, \qquad \Psi_i = \eps_{i j} \Psi^j~, \qquad \eps^{\1\2} = \eps_{\2\1} = 1~.
\end{align}

We employ a Weyl basis for the gamma matrices so that
an eight-component Dirac spinor $\Psi$ decomposes into a four-component
left-handed Weyl spinor $\psi^\alpha$ and a four-component right-handed spinor $\chi_\alpha$
so that
\begin{align}\label{eq:ChiralDecomp1}
\Psi =
\begin{pmatrix}
\psi^\alpha \\
\chi_\alpha
\end{pmatrix}~, \qquad
\Gamma_* =
\begin{pmatrix}
\delta^\a{}_\b & 0 \\
0 & -\delta_\a{}^\b
\end{pmatrix}~, \qquad \alpha=1,\cdots, 4~.
\end{align}
The spinors $\psi^\alpha$ and $\chi_\alpha$ are valued in the
two inequivalent fundamental representations 
of $\mathfrak{su}^*(4) \cong \mathfrak{so}(5,1)$.
We further take
\begin{align}
\Gamma^a =
\begin{pmatrix}
0 & (\tilde\gamma^a)^{\alpha\beta} \\
(\gamma^a)_{\alpha\beta} & 0 
\end{pmatrix}~,\qquad
C =
\begin{pmatrix}
0 & \delta_\alpha{}^\beta \\
\delta^\alpha{}_\beta & 0
\end{pmatrix}~.
\end{align}
The Pauli-type $4 \times 4$ matrices $(\gamma^a)_{\alpha\beta}$ 
and $(\tilde\gamma^a)^{\alpha\beta}$ are antisymmetric and related by
\begin{align}
(\tilde\gamma^a)^{\alpha\beta} = \frac{1}{2} \veps^{\a\b\g\d} (\gamma^a)_{\g\d}~, \qquad
(\gamma^a)^* = \tilde\gamma_a~,
\end{align}
where $\veps^{\a\b\g\d}$ is the canonical antisymmetric symbol of $\mathfrak{su}^*(4)$.
They obey
\bsubeq
\bea (\g^a)_{\a\b} (\tilde{\g}^b)^{\b\g}
+ (\g^b)_{\a\b} (\tilde{\g}^a)^{\b\g} &=& - 2 \eta^{ab} \d^\g_\a \ , \\
(\tilde{\g}^a)^{\a\b} (\g^b)_{\b\g}
+ (\tilde{\g}^b)^{\a\b} (\g^a)_{\b\g} 
&=& - 2 \eta^{ab} \d^\a_\g \ ,
\eea
\esubeq
and as a consequence of \eqref{eq:GammaReality},
\begin{align}
(\gamma^a)_{\alpha\beta} = B_{\alpha}{}^\dgamma B_\beta{}^{\ddelta}
	\big((\gamma^a)_{\gamma \delta}\big)^*~, \quad
(\tilde\gamma^a)^{\alpha\beta} = B^{\alpha}{}_\dgamma B^\beta{}_{\ddelta}
	\big((\tilde\gamma^a)^{\gamma \delta}\big)^*~, \quad
B =
\begin{pmatrix}
B^{\alpha}{}_\dbeta & 0 \\
0 & B_{\alpha}{}^\dbeta
\end{pmatrix}~.
\label{eq:PauliReality}
\end{align}
A dotted index denotes the complex conjugate representation in $\mathfrak{su}^*(4)$.
It is natural to use the $B$ matrix to define bar conjugation on a
four component spinor via
\begin{align}
\bar\psi^\alpha = B^\alpha{}_\dbeta (\psi^\beta)^*~, \qquad
\bar \chi_\alpha = B_\alpha{}^\dbeta (\chi_\beta)^*~,
\end{align}
with the obvious extension to any object with multiple spinor indices.
For example, $\overline{(\gamma^a)_{\alpha\beta}} = (\gamma^a)_{\alpha\beta}$
using \eqref{eq:PauliReality} and similarly for $\tilde\gamma^a$.
Note that $\overline{\overline{\psi^\a}} = -\psi^\a$ and similarly for any object with an
odd number of spinor indices as a consequence of $B^* B = -\bm{1}$.
A symplectic Majorana spinor $\Psi_i$, decomposed as in \eqref{eq:ChiralDecomp1}
and obeying \eqref{eq:SympMaj}, has Weyl components that obey
\begin{align}\label{eq:SympMaj4c}
\overline{\psi^{\alpha i}} = \psi^\alpha_{i}~, \qquad
\overline{\chi_{\alpha i}} = \chi_\alpha^{i}~.
\end{align}
The Grassmann coordinates $\q^\alpha_i$ and the parameters $\eta_\alpha^i$ of $S$-supersymmetry
are both symplectic Majorana-Weyl using this definition. 

We define the antisymmetric products of two or three Pauli-type matrices as
\bsubeq
\begin{alignat}{2}
\g_{ab} &:= \g_{[a} \tilde{\g}_{b]} := \hf (\g_a \tilde{\g}_b - \g_b \tilde{\g}_a) \ , &\quad
\tilde{\g}_{ab} &:= \tilde{\g}_{[a} \g_{b]}  = -(\g_{ab})^T\ , \\
\g_{abc} &:= \g_{[a} \tilde{\g}_b \g_{c]} \ , &\quad \tilde{\g}_{abc} &:= \tilde{\g}_{[a} \g_b \tilde{\g}_{c]} \ .
\end{alignat}
\esubeq
Note that $\g_{ab}$ and $\tilde\g_{ab}$ are traceless, whereas $\g_{abc}$ and
$\tilde\g_{abc}$ are symmetric. Further antisymmetric products obey
\begin{subequations}
\begin{alignat}{2}
\g_{abc} &= -\frac{1}{3!} \eps_{abcdef} \g^{def}~, &\qquad
\tilde \g_{abc} &= \frac{1}{3!} \eps_{abcdef} \tilde \g^{def}~, \\
\g_{abcd} &= \frac{1}{2} \eps_{abcdef} \g^{ef}~, &\qquad
\tilde \g_{abcd} &= -\frac{1}{2} \eps_{abcdef} \tilde \g^{ef}~, \\
\g_{abcde} &= \eps_{abcdef} \g^f~, &\qquad
\tilde \g_{abcde} &= -\eps_{abcdef} \tilde \g^f~, \\
\g_{abcdef} &= -\eps_{abcdef}~, &\qquad
\tilde \g_{abcdef} &= \eps_{abcdef}~.
\end{alignat}
\end{subequations}

Making use of the completeness relations
\begin{subequations}
\begin{align}
(\gamma^a)_{\a\b} (\tilde\gamma_{a})^{\g\d} &= 4\, \delta_{[\a}{}^\g \delta_{\b]}{}^{\d}~, \\
(\gamma^{ab})_\a{}^\b (\gamma_{ab})_\g{}^\d &= - 8\,\delta_{\a}{}^\d \delta_{\g}{}^{\b}
	+ 2\, \delta_{\a}{}^\b \delta_{\g}{}^{\d}~, \\
(\gamma^{abc})_{\a\b} (\tilde\gamma_{abc})^{\g\d} &= 48\, \delta_{(\a}{}^\g \delta_{\b)}{}^{\d}~, \\
(\gamma^{abc})_{\a\b} (\tilde\gamma_{abc})_{\g\d} &= (\gamma^{abc})^{\a\b} (\tilde\gamma_{abc})^{\g\d} = 0~,
\end{align}
\end{subequations}
it is straightforward to establish natural isomorphisms between tensors of $\mathfrak{so}(5,1)$
and matrix representations of $\mathfrak{su}^*(4)$.
Vectors $V^a$ and antisymmetric matrices $V_{\a\b} = - V_{\b\a}$ 
are related by
\be 
V_{\a\b} := (\g^a)_{\a\b} V_a \quad \Longleftrightarrow  \quad V_a = \frac{1}{4} (\tilde{\g}_a)^{\a\b} V_{\a\b} \ .
\ee
Antisymmetric rank-two tensors $F_{ab}$ are related to traceless matrices $F_\a{}^\b$ 
via
\be F_\a{}^\b := - \frac{1}{4} (\g^{ab})_\a{}^\b F_{ab} \ , \quad F_\a{}^\a = 0 \quad
\Longleftrightarrow 
\quad F_{ab} = \hf (\g_{ab})_\b{}^\a F_\a{}^\b = - F_{ba} \ .
\ee
Self-dual and anti-self-dual rank-three antisymmetric tensors $T^{(\pm)}_{abc}$,
\be \frac{1}{3!} \eps^{abcdef} T_{def}^{(\pm)} = \pm T^{(\pm)abc} \ ,
\ee
are related to symmetric matrices $T_{\a\b}$ and $T^{\a\b}$ 
via
\bsubeq
\bea
T_{\a\b} &:=& \frac{1}{3!} (\g^{abc})_{\a\b} T_{abc} = T_{\b\a} \quad \Longleftrightarrow \quad 
T_{abc}^{(+)} = \frac{1}{8} (\tilde{\g}_{abc})^{\a\b} T_{\a\b} \ , \\
T^{\a\b} &:=& \frac{1}{3!} (\tilde{\g}^{abc})^{\a\b} T_{abc} = T^{\b\a} \quad 
\Longleftrightarrow \quad
T^{(-)}_{abc} = \frac{1}{8} (\g_{abc})_{\a\b} T^{\a\b} \ .
\eea
\esubeq
Further irreducible representations of the Lorentz group take particularly
simple forms when written with spinor indices. For example, a gamma-traceless
left-handed spinor two-form $\Psi_{ab}{}^{\g}$ is related to a symmetric traceless
$\Psi_\a{}^{\b\g}$, 
\begin{align}
\Psi_\a{}^{\b \g} &:= - \frac{1}{4} (\g^{ab})_\a{}^\b \Psi_{ab}{}^\g
	= \Psi_\a{}^{\g \b}\ , \quad \Psi_\a{}^{\a\g} = 0 \quad
\Longleftrightarrow \eol
\Psi_{ab}{}^\g &= \hf (\g_{ab})_\b{}^\a \Psi_\a{}^{\b \g} \ ,\quad
(\gamma^b)_{\d\g} \Psi_{ab}{}^\g = 0~,
\end{align}
and a rank-four tensor $C_{abcd}$ with the symmetries of the Weyl
tensor is related to a symmetric traceless $C_{\a\g}{}^{\b\d}$ 
via
\begin{align}
C_{\a\g}{}^{\b\d} &:= \frac{1}{16} (\g^{ab})_\a{}^\b (\g^{cd})_\g{}^\d \, C_{abcd}
	= C_{(\a\g)}{}^{(\b\d)}~, \qquad C_{\a\g}{}^{\b\g} = 0
	\quad \Longleftrightarrow \eol
C_{abcd} &= \frac{1}{4} (\g_{ab})_\b{}^\a (\g_{cd})_\d{}^\g\, C_{\a\g}{}^{\b\d}
	= C_{[cd] [ab]}~, \qquad
	C_{[abc]d} = 0~.
\end{align}


\section{The conformal Killing supervector fields of $\mathds R^{6|8}$}
\label{KVF}

Simple Minkowski superspace in six dimensions, $\mathds R^{6|8}$,
 is parametrized by coordinates $z^{A} = (x^a , \q^{\a}_i)$. The flat covariant 
derivatives
$D_{A} = (\partial_a , D_{\a}^i)$
\bea
\pa_a:=\frac{\pa}{\pa x^a}~,~~~~~~
D_{\a}^i
:=
\frac{\pa}{\pa\q^{\a}_i}
-\ri(\g^a)_{\a\b} \q^{\b i} \pa_a
~,
\eea
 satisfy the algebra:
\bea
\{ D_{\a}^i , D_{\b}^j \} = - 2 \ri \eps^{ij} (\g^a)_{\a\b} \partial_a 
~, 
~~~~~~
{[}\partial_a , D_{\b}^j{]} = 0 
~,~~~~~~
{[}\pa_a,\pa_b{]}=0
~.
\eea

The conformal Killing supervector fields
\be \xi = \bar{\xi} = \xi^a \partial_a + \xi^\a_i D_\a^i
\ee
may be defined to satisfy
\be [\xi , D_\a^i ] = - (D_\a^i \xi^\b_j) D_\b^j \ , \label{2.22}
\ee
which implies the fundamental equation
\be D_\a^i \xi_a = -2 \ri (\g_a)_{\a\b} \xi^{\b i} \ . \label{2.23}
\ee
From eq. \eqref{2.23} one finds
\be \eps^{ij} (\g^b)_{\a\b} \partial_b \xi_a = (\g_a)_{\a\g} D_\b^j \xi^{\g i} + (\g_a)_{\b\g} D_\a^i \xi^{\g j} \ ,
\ee
which gives us the equation for a conformal Killing vector field,
\be \partial_{(a} \xi_{b)}= \frac{1}{6} \eta_{a b} \partial^c \xi_c \ , \label{BosonicKilling}
\ee
as well as the following useful identities:
\bsubeq
\bea
D_\a^{(i} \xi^{\b j)} &=& \frac{1}{4} \d^\b_\a D_\g^{(i} \xi^{\g j)} \ , \\
D_\g^k \xi^\g_k &=& \frac{2}{3} \partial^a \xi_a \ , \\
D_\a^k \xi^\b_k 
- \frac{1}{4} \d_\a^\b D_\g^k \xi_k^\g 
&=& - \hf (\g^{ab})_\a{}^\b \partial_{a} \xi_{b} \ .
\eea
\esubeq

The conformal Killing supervector field acts on the spinor covariant derivatives as follows
\be [\xi , D_\a^i] = - \omega_\a{}^\b D_\b^i + \L^{ij} D_{\a j} - \hf \s D_\a^i  \ ,
\ee
where the parameters $\omega_\a{}^\b$, $\s$ and $\L^{ij}$ are given by the following expressions:
\bsubeq \label{paraMeteRs}
\begin{align}
\omega_\a{}^\b &:= - \frac{1}{4} (\g^{ab})_\a{}^\b \partial_a \xi_b \ ,\\
\s &:= \frac{1}{4} D_\g^k \xi_k^\g = - \frac{1}{6} \partial^a \xi_a \ , \\
\L^{ij} &:= \frac{1}{4} D_\g^{(i} \xi^{\g j)} \ .
\end{align}
\esubeq
Using eq. \eqref{BosonicKilling} one finds that the parameters \eqref{paraMeteRs} satisfy
\bsubeq
\begin{align}
\partial_{a} \omega_{b c} &= - 2 \eta_{a [b} \partial_{c]} \s \ , \\
\partial_{a} \partial_{b} \xi_{c} &= \eta_{ab} \partial_c \s - 2 \eta_{c (a} \partial_{b )} \s \ ,
\end{align}
\esubeq
while using eq. \eqref{2.23} one finds
\bsubeq
\begin{align}
D_\g^k \o_\a{}^\b &= 
2 \d_\g^\b D_\a^k \s - \hf \d^\b_\a D_\g^k \s
\ , \\
D_\a^i \L^{jk} &= - 4 \eps^{i(j} D_\a^{k)} \s \ ,
\end{align}
\esubeq
where $\s$ obeys
\be D_\a^i D_\b^j \s = -  \ri \eps^{ij} \partial_{\a\b} \s \ , \quad \partial_a D_\b^j \s = 0 \ . 
\ee
Finally, one can verify that the following holds
\be \partial_a \xi^\g_k = \frac{\ri}{2} (\tilde{\g}_a)^{\b\g} D_{\b k} \s \ .
\ee

The above results tell us that we can parametrize superconformal Killing vectors as
\be \xi \equiv \xi(\l(P)^a , \l(Q)^\a_i \ , \l(M)_{ab} , \l(J)^{ij} , \l({\mathbb D}) , \l(K)_a , \l(S)_\a^i) \ ,
\ee
where we have defined the parameters
\bsubeq
\begin{align} \l(P)^a &:= \xi^a |_{x = \theta = 0} \ , \quad \l(Q)^\a_i = \xi^\a_i|_{x = \theta = 0} \ , \\
\l(M)_{ab} &:= \omega_{ab}|_{x = \theta = 0} \ , \quad \l({\mathbb D}) := \s|_{x = \theta = 0} \ , \quad \l(J)^{ij} = \L^{ij}|_{x = \theta = 0} \ , \\
\l(K)_a &:= \hf \partial_a \s |_{x = \theta = 0} \ , \quad \l(S)_{\a}^i := \eta_\a^i|_{x = \theta = 0} \ ,
\end{align}
\esubeq
and we have introduced
\be \eta_\a^i := \hf D_\a^i \s \ .
\ee
The commutator of two superconformal Killing vectors,
\be \xi = \xi(\l(P)^a , \l(Q)^\a_i \ , \l(M)_{ab} , \l(J)^{ij}, \l({\mathbb D}) , \l(K)^a , \l(S)_\a^i)
\ee
and
\be \tilde{\xi} = \xi(\tilde{\l}(P)^a , \tilde{\l}(Q)^\a_i , \tilde{\l}(M)_{ab} , \tilde{\l}(J)^{ij} , \tilde{\l}({\mathbb D}) , \tilde{\l}(K)^a , \tilde{\l}(S)_\a^i) \ ,
\ee
is another superconformal Killing vector given by
\begin{align} [\xi , \tilde{\xi}] &= (\xi^a \partial_a \tilde{\xi}^b 
- \tilde{\xi}^a \partial_a \xi^b 
+ \xi^\a_i D_\a^i \tilde{\xi}^b 
- \tilde{\xi}^\a_i D_\a^i \xi^b 
+ 2 \ri \xi^\a_k \tilde{\xi}^{\b k} (\g^b)_{\a\b}) \partial_b 
\non\\
& + (\xi^a \partial_a \tilde{\xi}^\b_j
- \tilde{\xi}^a \partial_a \xi^\b_j 
+ \xi^\a_i D_\a^i \tilde{\xi}^\b_j - \tilde{\xi}^\a_i D_\a^i \xi^\b_j) D_\b^j \non\\
&= \Big(\xi^a \tilde{\omega}_a{}^b 
+ \xi^b \tilde{\s}
- \tilde{\xi}^a \omega_a{}^b 
- \tilde{\xi}^b \s
- 2 \ri \xi^\a_k \tilde{\xi}^{\b k} (\g^b)_{\a \b} \Big) \partial_b \non\\
& + \Big( - \ri \xi^a (\tilde{\g}_a)^{\b\g} \tilde{\eta}_{\g j}
+ \hf \xi^\b_j \tilde{\s}
 - \xi^\a_j \tilde{\omega}_\a{}^\b 
  + \xi^\b_i \tilde{\L}^i{}_j \non\\
&\qquad 
+ \ri \tilde{\xi}^a (\tilde{\g}_a)^{\b\g} \eta_{\g j} 
- \hf \tilde{\xi}^\b_j \s
+ \tilde{\xi}^\a_j \omega_\a{}^\b
- \tilde{\xi}^\b_i \L^i{}_j \Big) D_\b^j
\non\\
&\equiv 
\xi(\hat{\l}(P)^a , \hat{\l}(Q)^\a_i , \hat{\l}(M)_{ab} , \hat{\l}(J)^{ij} , \hat{\l}({\mathbb D}) , \hat{\l}(K)^a , \hat{\l}(S)_\a^i) \ ,
\end{align}
where
\bsubeq \label{compsXi}
\begin{align}
\hat{\l}^{a}(P) &:=  \l(P)^b \tilde{\l}(M)_b{}^a 
+ \l(P)^a \tilde{\l}({\mathbb D}) 
- 2 \ri \l(Q)^\a_k \tilde{\l}(Q)^{\b k} (\g^a)_{\a\b} \non\\
&\qquad- \tilde{\l}(P)^b \l(M)_b{}^a 
- \tilde{\l}(P)^a \l({\mathbb D})  \ , \\
\hat{\l}^\a_i(Q) &:= - \ri (\tilde{\g}_a)^{\a\b} \l(P)^a \tilde{\l}(S)_{\b i} 
- \l(Q)^\b_i \tilde{\l}(M)_\b{}^\a 
+ \hf \l(Q)^\a_i \tilde{\l}({\mathbb D}) 
+ \l(Q)^\a_j \tilde{\l}(J)^j{}_i \non\\
&\quad 
+ \ri (\tilde{\g}_a)^{\a\b} \tilde{\l}(P)^a \l(S)_{\b i} 
+ \tilde{\l}(Q)^\b_i \l(M)_\b{}^\a 
- \hf \tilde{\l}(Q)^\a_i \l({\mathbb D}) 
- \tilde{\l}(Q)^\a_j \l(J)^j{}_i \ , \\
\hat{\l}(M)_{ab} &:= 2 \l(M)_{[a}{}^c \tilde{\l}(M)_{b] c} 
- 4 \l(P)_{[\ha} \tilde{\l}(K)_{\hb]} 
+ 4 \tilde{\l}(P)_{[\ha} \l(K)_{\hb]} \non\\
&\qquad+ 2 (\g_{ab})_\a{}^\b \l(Q)^\a_k \tilde{\l}(S)_\b^k 
- 2 (\g_{ab})_\a{}^\b \tilde{\l}(Q)^\a_k \l(S)_\b^k \ , \\
\hat{\l}(J)^{ij} &:=
2 \l(J)_k{}^{(i} \tilde{\l}(J)^{j) k}
- 8 \l(Q)^{\g (i} \tilde{\l}(S)_\g^{j)}
+ 8 \tilde{\l}(Q)^{\g (i} \l(S)_\g^{j)} \ ,
\\
\hat{\l}({\mathbb D}) &:= 2 \l(P)^a \tilde{\l}(K)_a 
- 2 \tilde{\l}(P)^a \l(K)_a 
+ 2 \l(S)_\a^i \tilde{\l}(Q)^\a_i 
- 2 \tilde{\l}(S)_\a^i \l(Q)^\a_i \ , \\
\hat{\l}(K)^a &:= \l(M)^{a b} \tilde{\l}(K)_b 
+ \l({\mathbb D}) \tilde{\l}(K)^a 
+ 2 \ri (\tilde{\g}_a)^{\a\b} \tilde{\l}(S)_\a^k \l(S)_{\b k} \non\\
&\qquad- \tilde{\l}(M)^{a b} \l(K)_\hb - \tilde{\l}({\mathbb D}) \l(K)^a  \ , \\
\hat{\l}(S)_\a^i &:= \ri (\g_a)_{\a\b} \l(K)^a \tilde{\l}(Q)^{\b i} 
+ \l(S)_\b^i \tilde{\l}(M)_\a{}^\b 
- \hf \l(S)_\a^i \tilde{\l}({\mathbb D}) 
- \l(S)_\a^j \tilde{\l}(J)^i{}_j \non\\
&\quad 
- \ri (\g_a)_{\a\b} \tilde{\l}(K)^a \l(Q)^{\b i} 
-  \tilde{\l}(S)_\b^i \l(M)_\a{}^\b 
+ \hf \tilde{\l}(S)_\a^i \l({\mathbb D}) 
+ \tilde{\l}(S)_\a^j \l(J)^i{}_j  \ .
\end{align}
\esubeq

Representing the superconformal Killing vectors as
\bea 
\xi &=& \l(P)^a P_a + \l(Q)^\a_i Q_\a^i + \hf \l(M)^{ab} M_{ab} 
+ \l(J)^{ij} J_{ij} + \l({\mathbb D}) {\mathbb D} \non\\
&&+ \l(K)^a K_a + \l(S)_\a^i S^\a_i
\eea
and comparing eq. \eqref{compsXi} to the commutator
\be [\xi , \tilde{\xi}] = - \tilde{\l}^{\underline{b}} \l^{\underline{a}} [X_{\underline{a}} , X_{\underline{b}}\}
\ee
gives the superconformal algebra.


\section{The Yang-Mills multiplet in conformal superspace} \label{YMmultiplet}

To describe a non-abelian vector multiplet, the covariant derivative $\nabla = E^A \nabla_A $ 
has to be replaced with a gauge covariant one, 
\bea
\bm \nabla = E^A \bm \nabla_A \ , \quad {\bm\nabla}_A := \nabla_A 
- \ri \bm V_A
~.
\label{SYM-derivatives}
\eea
Here the  gauge connection one-form $\bm V = E^A \bm V_A$ 
takes its values in the Lie algebra 
of the (unitary) Yang-Mills gauge group, $G_{\rm YM}$, with its (Hermitian) generators 
commuting with all the generators of the superconformal algebra. 
The algebra of the gauge covariant derivatives  is
\bea
[{\bm \nabla}_A, {\bm \nabla}_B \} 
&=&
 -\scT_{AB}{}^C{\bm \nabla}_C
-\hf \sRM_{AB}{}^{cd} M_{cd}
-\sRJ_{AB}{}^{kl} J_{kl}
- \sRD_{AB} \mathbb D 
\non\\
&&
 - \sRS_{AB}{}^\g_k S_\g^k
	- \sRK_{AB}{}^c K_c
	- \ri \bm F_{AB} \ ,
\eea
where the torsion and curvatures are those of conformal superspace 
but with $\bm F_{AB}$ corresponding 
to the gauge covariant field strength two-form 
$\bm F = \hf E^B \wedge E^A \bm F_{AB}$.
The field strength $\bm F_{AB}$ 
satisfies the Bianchi identity
\be \bm \nabla \bm F = 0  \quad \Longleftrightarrow \quad
\bm \nabla_{[A} \bm F_{BC\}} 
+ \scT_{[AB}{}^D \bm F_{|D| C\}} = 0
~. \label{FBI45}
\ee
The Yang-Mills gauge transformation acts on the gauge covariant 
derivatives $\bm \nabla_A$ and a matter  superfield $U$ (transforming 
in some representation of the gauge group) 
as
\be 
\bm \nabla_A ~\rightarrow~ \re^{\ri  \bm  \t} \bm \nabla_A \re^{- \ri \bm \t } , 
\qquad  U~\rightarrow~ U' = \re^{\ri  \bm  \t} U~, 
\qquad \bm \t^\dag = \bm \t \ ,
\label{2.2}
\ee
where the Hermitian gauge parameter ${\bm \t} (z)$ takes its values in the Lie algebra 
of $G_{\rm YM}$. 
This implies that the gauge one-form and the field strength transform as follows:
\bea 
\bm V ~\rightarrow ~\re^{\ri \bm \t} \bm V \re^{-\ri \bm \t} 
+ \ri \,\re^{\ri \bm \t} \rd \, \re^{- \ri \bm \t} \ , \qquad 
\bm F ~\rightarrow ~ \re^{\ri \bm \t} \bm F \re^{- \ri \bm \t} \ .
\eea

Some components of the field strength have to be constrained 
in order to describe an irreducible multiplet. 
The constraints are (see e.g. \cite{HST})
\bsubeq
\bea 
\bm F_\a^i{}_\b^j = 0 \ , \quad \bm F_a{}_\b^j = (\g_a)_{\a\b} \bm W^{\b i} \ , \label{YMsuperformFa}
\eea
where $\bm W^{\a i}$ is a conformal primary of dimension 3/2,
 $S^\g_k \bm W^{\a i} =0$ and
 $\bbD \bm W^{\a i}= \frac{3}{2} \bm W^{\a i}$. The Bianchi identity \eqref{FBI45} together with the constraints \eqref{YMsuperformFa} 
 fix the remaining component of the field strength to be
\bea
\bm F_{ab} = - \frac{\ri}{8} (\g_{ab})_\b{}^\a \bm \nabla_\a^k \bm W^\b_k
\eea
\esubeq
and constrain $\bm W^{\a i}$ to obey the differential constraints
\bea 
\bm \nabla_\g^k \bm W^\g_k = 0 \ , \quad \bm \nabla_\a^{(i} \bm W^{\b j)}
= \frac{1}{4} \d_\a^\b \bm \nabla_\g^{(i} \bm W^{\g j)}
\label{vector-Bianchies} \ .
\eea

It is helpful to introduce the following descendant superfield:
\begin{align}
\bm X^{ij} := \frac{\ri}{4} \bm\nabla^{(i}_\g \bm W^{\g j)} \ .
\label{C.8}
\end{align}
The superfield $\bm W^{\a i}$ and $\bm X^{ij}$, together with
\be 
\bm F_\a{}^\b = - \frac{\ri}{4} \Big( \bm \nabla_\a^k \bm W^\b_k - \frac{1}{4} \d_\a^\b \bm \nabla_\g^k \bm W^\g_k \Big)
= - \frac{\ri}{4} \bm \nabla_\a^k \bm W^\b_k \ ,
\ee
satisfy the following useful identities:
\bsubeq \label{VMIdentities}
\bea
\bm\nabla_\a^i \bm W^{\b j}
&=& 
- \ri \d_\a^\b \bm X^{ij} - 2 \ri \eps^{ij} \bm F_\a{}^\b \ , \\
\bm \nabla_\a^i \bm F_\b{}^\g
&=&
- \bm \nabla_{\a\b} \bm W^{\g i}
- \d^\g_\a \bm \nabla_{\b\d} \bm W^{\d i}
+ \frac{1}{2} \d_\b^\g \bm \nabla_{\a \d} \bm W^{\d i}
\ , \\
\bm\nabla_\a^i \bm X^{jk}
&=&
2 \eps^{i(j} \bm \nabla_{\a\b} \bm W^{\b k)}
\ .
\eea
\esubeq
The $S$-supersymmetry generator acts on these descendants as
\begin{align}
S^\g_k \bm F_\a{}^\b = - 4 \ri \d^\g_\a \bm W^\b_k
+ \ri \d^\b_\a \bm W^\g_k \ , \qquad
S^\g_k \bm X^{ij} = - 4 \ri \d_k^{(i} \bm W^{\g j)} \ .
\end{align}


\begin{footnotesize}

\end{footnotesize}


\begin{thebibliography}{66}

\bibitem{Nahm} 
  W.~Nahm,
  ``Supersymmetries and their representations,''
  Nucl.\ Phys.\ B {\bf 135}, 149 (1978).
  
  \bibitem{ISZ05} 
  E.~A.~Ivanov, A.~V.~Smilga and B.~M.~Zupnik,
  ``Renormalizable supersymmetric gauge theory in six dimensions,''
  Nucl.\ Phys.\ B {\bf 726}, 131 (2005)
  [hep-th/0505082].

\bibitem{SW}
  N.~Seiberg and E.~Witten,
  ``Comments on string dynamics in six-dimensions,''
  Nucl.\ Phys.\ B {\bf 471}, 121 (1996)
  [hep-th/9603003].
  
\bibitem{Seiberg} 
  N.~Seiberg,
  ``Nontrivial fixed points of the renormalization group in six-dimensions,''
  Phys.\ Lett.\ B {\bf 390}, 169 (1997)
  [hep-th/9609161].

\bibitem{Witten1} 
  E.~Witten,
  ``Some comments on string dynamics,''
  [hep-th/9507121].

\bibitem{Strominger} 
  A.~Strominger,
  ``Open p-branes,''
  Phys.\ Lett.\ B {\bf 383}, 44 (1996)
  [hep-th/9512059].

\bibitem{Witten2} 
  E.~Witten,
  ``Five-branes and M theory on an orbifold,''
  Nucl.\ Phys.\ B {\bf 463}, 383 (1996)
  [hep-th/9512219].

\bibitem{GH} 
  O.~J.~Ganor and A.~Hanany,
  ``Small E(8) instantons and tensionless noncritical strings,''
  Nucl.\ Phys.\ B {\bf 474}, 122 (1996)
  [hep-th/9602120].

\bibitem{BI} 
  J.~D.~Blum and K.~A.~Intriligator,
  ``New phases of string theory and 6-D RG fixed points via branes at orbifold  
  singularities,''
  Nucl.\ Phys.\ B {\bf 506}, 199 (1997)
  [hep-th/9705044].

\bibitem{HZ} 
  A.~Hanany and A.~Zaffaroni,
  ``Branes and six-dimensional supersymmetric theories,''
  Nucl.\ Phys.\ B {\bf 529}, 180 (1998)
  [hep-th/9712145].

\bibitem{HMV} 
  J.~J.~Heckman, D.~R.~Morrison and C.~Vafa,
  ``On the classification of 6D SCFTs and generalized ADE orbifolds,''
  JHEP {\bf 1405}, 028 (2014)
  Erratum: [JHEP {\bf 1506}, 017 (2015)]
  [arXiv:1312.5746 [hep-th]].

\bibitem{DZHTV} 
  M.~Del Zotto, J.~J.~Heckman, A.~Tomasiello and C.~Vafa,
  ``6d conformal matter,''
  JHEP {\bf 1502}, 054 (2015)
  [arXiv:1407.6359 [hep-th]].

\bibitem{HMRV} 
  J.~J.~Heckman, D.~R.~Morrison, T.~Rudelius and C.~Vafa,
  ``Atomic classification of 6D SCFTs,''
  Fortsch.\ Phys.\  {\bf 63}, 468 (2015)
  [arXiv:1502.05405 [hep-th]].

\bibitem{DS}
S.~Deser and A.~Schwimmer,
  ``Geometric classification of conformal anomalies in arbitrary dimensions,''
  Phys.\ Lett.\ B {\bf 309}, 279 (1993)
  [hep-th/9302047].

\bibitem{WZ} 
  J.~Wess and B.~Zumino,
  ``Consequences of anomalous Ward identities,''
  Phys.\ Lett.\ B {\bf 37}, 95 (1971).

\bibitem{BPT}
  L.~Bonora, P.~Pasti and M.~Tonin,
``Cohomologies and anomalies in supersymmetric theories,''
  Nucl.\ Phys.\ B {\bf 252}, 458 (1985).

\bibitem{BK86}
  I.~L.~Buchbinder and S.~M.~Kuzenko,
  ``Matter superfields in external supergravity: Green functions, effective action and superconformal anomalies,''
  Nucl.\ Phys.\ B {\bf 274}, 653 (1986).

\bibitem{K_N=2SW} 
  S.~M.~Kuzenko,
 ``Super-Weyl anomalies in N=2 supergravity and (non)local effective actions,''
  JHEP {\bf 1310}, 151 (2013)
  [arXiv:1307.7586 [hep-th]].

\bibitem{CDI2} 
  C.~Cordova, T.~T.~Dumitrescu and K.~Intriligator,
  ``Deformations of superconformal theories,''
  arXiv:1602.01217 [hep-th].

\bibitem{LL} 
  J.~Louis and S.~L\"ust,
  ``Supersymmetric AdS$_{7}$ backgrounds in half-maximal supergravity 
  and marginal operators of (1, 0) SCFTs,''
  JHEP {\bf 1510}, 120 (2015)
  [arXiv:1506.08040 [hep-th]].

\bibitem{FHMM} 
  D.~Freed, J.~A.~Harvey, R.~Minasian and G.~W.~Moore,
  ``Gravitational anomaly cancellation for M theory five-branes,''
  Adv.\ Theor.\ Math.\ Phys.\  {\bf 2}, 601 (1998)
  [hep-th/9803205].

\bibitem{HS} 
  M.~Henningson and K.~Skenderis,
  ``The holographic Weyl anomaly,''
  JHEP {\bf 9807}, 023 (1998)
  [hep-th/9806087].

\bibitem{HMM} 
  J.~A.~Harvey, R.~Minasian and G.~W.~Moore,
  ``NonAbelian tensor multiplet anomalies,''
  JHEP {\bf 9809}, 004 (1998)
  [hep-th/9808060].

\bibitem{Intriligator} 
  K.~A.~Intriligator,
  ``Anomaly matching and a Hopf-Wess-Zumino term in 6d, N=(2,0) field theories,''
  Nucl.\ Phys.\ B {\bf 581}, 257 (2000)
  [hep-th/0001205].

\bibitem{Yi} 
  P.~Yi,
  ``Anomaly of (2,0) theories,''
  Phys.\ Rev.\ D {\bf 64}, 106006 (2001)
  [hep-th/0106165].

\bibitem{OSTY} 
  K.~Ohmori, H.~Shimizu, Y.~Tachikawa and K.~Yonekura,
  ``Anomaly polynomial of general 6d SCFTs,''
  PTEP {\bf 2014}, no. 10, 103B07 (2014)
  [arXiv:1408.5572 [hep-th]].

\bibitem{CDY} 
  C.~Cordova, T.~T.~Dumitrescu and X.~Yin,
  ``Higher derivative terms, toroidal compactification, and Weyl anomalies in six-dimensional (2,0) theories,''
  arXiv:1505.03850 [hep-th].

\bibitem{FT85} 
  E.~S.~Fradkin and A.~A.~Tseytlin,
  ``Conformal supergravity,''
  Phys.\ Rept.\  {\bf 119}, 233 (1985).

\bibitem{BSVanP} 
  E.~Bergshoeff, E.~Sezgin and A.~Van Proeyen,
  ``Superconformal tensor calculus and matter couplings in six dimensions,''
  Nucl.\ Phys.\ B {\bf 264}, 653 (1986)
  Erratum: [Nucl.\ Phys.\ B {\bf 598}, 667 (2001)].

\bibitem{deWvHVP1}
 B.~de Wit, J.~W.~van Holten and A.~Van Proeyen,
 ``Transformation rules of $N=2$ supergravity multiplets,''
Nucl.\ Phys.\  B {\bf 167}, 186 (1980).
  
\bibitem{deRvHdeWVP} 
  M.~de Roo, B.~de Wit,   J.~W.~van Holten and A.~Van Proeyen,
  ``Chiral superfields in $N=2$ supergravity,''
  Nucl.\ Phys.\ B {\bf 173}, 175 (1980).  

\bibitem{deWvHVP2}
  B.~de Wit, J.~W.~van Holten and A.~Van Proeyen,
  ``Structure of $N=2$ supergravity,''
  Nucl.\ Phys.\ B {\bf 184}, 77 (1981)
  Erratum: [Nucl.\ Phys.\ B {\bf 222}, 516 (1983)].

\bibitem{deWPVP} 
  B.~de Wit, R.~Philippe and A.~Van Proeyen,
  ``The improved tensor multiplet in $N=2$ supergravity,''
  Nucl.\ Phys.\ B {\bf 219}, 143 (1983).

\bibitem{deWLPSVP} 
  B.~de Wit, P.~G.~Lauwers, R.~Philippe, S.~Q.~Su and A.~Van Proeyen,
  ``Gauge and matter fields coupled to $N=2$ supergravity,''
  Phys.\ Lett.\ B {\bf 134}, 37 (1984).

\bibitem{deWLVP}
B.~de Wit, P.~G.~Lauwers and A.~Van Proeyen,
``Lagrangians of $N=2$ supergravity-matter systems,''
Nucl.\ Phys.\  B {\bf 255}, 569 (1985).
  
\bibitem{FVP} 
  D.~Z.~Freedman and A.~Van Proeyen,
 {\it Supergravity}, 
   Cambridge, UK: Cambridge Univ. Pr. (2012) 607 p. 
  
\bibitem{BSS1} 
  E.~Bergshoeff, A.~Salam and E.~Sezgin,
  ``A supersymmetric $R^2$-action in six dimensions and torsion,''
  Phys.\ Lett.\ B {\bf 173}, 73 (1986).  
  
\bibitem{BSS2} 
  E.~Bergshoeff, A.~Salam and E.~Sezgin,
  ``Supersymmetric $R^2$ actions, conformal invariance and Lorentz Chern-Simons term in 6 and 10 dimensions,''
  Nucl.\ Phys.\ B {\bf 279}, 659 (1987).  
  
\bibitem{BR} 
  E.~Bergshoeff and M.~Rakowski,
  ``An off-shell superspace $R^2$-action in six dimensions,''
  Phys.\ Lett.\ B {\bf 191}, 399 (1987).  

  \bibitem{CVanP} 
  F.~Coomans and A.~Van Proeyen,
  ``Off-shell $N=(1,0)$, $D=6$ supergravity from superconformal methods,''
  JHEP {\bf 1102}, 049 (2011)
  Erratum: [JHEP {\bf 1201}, 119 (2012)]
  [arXiv:1101.2403 [hep-th]].

\bibitem{BCSVanP} 
  E.~Bergshoeff, F.~Coomans, E.~Sezgin and A.~Van Proeyen,
  ``Higher derivative extension of $6D$ chiral gauged supergravity,''
  JHEP {\bf 1207}, 011 (2012)
  [arXiv:1203.2975 [hep-th]].


\bibitem{NS2} 
  H.~Nishino and E.~Sezgin,
  ``New couplings of six-dimensional supergravity,''
  Nucl.\ Phys.\ B {\bf 505}, 497 (1997)
  [hep-th/9703075].
  
 \bibitem{FRS} 
  S.~Ferrara, F.~Riccioni and A.~Sagnotti,
  ``Tensor and vector multiplets in six-dimensional supergravity,''
  Nucl.\ Phys.\ B {\bf 519}, 115 (1998)
  [hep-th/9711059]. 
  
\bibitem{SalamS} 
  A.~Salam and E.~Sezgin,
 ``Chiral compactification on Minkowski ${}\times  S^2$ of 
 $N=2$ Einstein-Maxwell supergravity in six dimensions,''
  Phys.\ Lett.\ B {\bf 147}, 47 (1984).
    
\bibitem{NS1} 
H.~Nishino and E.~Sezgin,
``Matter and gauge couplings of N=2 supergravity in six dimensions,''
Phys.\ Lett.\ B {\bf 144}, 187 (1984). 
  
\bibitem{VanP} 
A.~Van Proeyen,
``Superconformal symmetry and higher-derivative Lagrangians,''
Springer Proc.\ Phys.\  {\bf 153}, 1 (2014)
[arXiv:1306.2169 [hep-th]].


\bibitem{Howe1}
P.~S.~Howe,
``A superspace approach to extended conformal supergravity,''
Phys.\ Lett.\  B {\bf 100}, 389 (1981).


\bibitem{Howe2}
P.~S.~Howe,
``Supergravity in superspace,''  Nucl.\ Phys.\  B {\bf 199}, 309 (1982).


\bibitem{KT-M08}
  S.~M.~Kuzenko and G.~Tartaglino-Mazzucchelli,
  ``Super-Weyl invariance in 5D supergravity,''
  JHEP {\bf 0804}, 032 (2008)
   [arXiv:0802.3953 [hep-th]].
   
\bibitem{LT-M12} 
  W.~D.~Linch III and G.~Tartaglino-Mazzucchelli,
  ``Six-dimensional supergravity and projective superfields,''
  JHEP {\bf 1208}, 075 (2012)
  [arXiv:1204.4195 [hep-th]].
   
\bibitem{HIPT}
P.~S.~Howe, J.~M.~Izquierdo, G.~Papadopoulos and P.~K.~Townsend,
``New supergravities with central charges and Killing spinors in 2+1 dimensions,''
Nucl.\ Phys.\  B {\bf 467}, 183 (1996)  [arXiv:hep-th/9505032].
    
\bibitem{KLT-M}
 S.~M.~Kuzenko, U.~Lindstr\"om and G.~Tartaglino-Mazzucchelli,
``Off-shell supergravity-matter couplings in three dimensions,''
JHEP {\bf 1103}, 120 (2011)  [arXiv:1101.4013 [hep-th]].

\bibitem{Butter4DN=1} 
  D.~Butter,
  ``N=1 conformal superspace in four dimensions,''
  Annals Phys.\  {\bf 325}, 1026 (2010)
  [arXiv:0906.4399 [hep-th]].
  
\bibitem{Butter4DN=2} 
  D.~Butter,
  ``N=2 conformal superspace in four dimensions,''
  JHEP {\bf 1110}, 030 (2011)
  [arXiv:1103.5914 [hep-th]].
  
\bibitem{BKNT-M1} 
  D.~Butter, S.~M.~Kuzenko, J.~Novak and G.~Tartaglino-Mazzucchelli,
  ``Conformal supergravity in three dimensions: New off-shell formulation,''
  JHEP {\bf 1309}, 072 (2013)
  [arXiv:1305.3132 [hep-th]].
  
\bibitem{BKNT-M5D} 
  D.~Butter, S.~M.~Kuzenko, J.~Novak and G.~Tartaglino-Mazzucchelli,
  ``Conformal supergravity in five dimensions: New approach and applications,''
  JHEP {\bf 1502}, 111 (2015)
  [arXiv:1410.8682 [hep-th]].


\bibitem{Galperin:1985zv} 
  A.~Galperin, E.~Ivanov, V.~Ogievetsky and E.~Sokatchev,
  ``Conformal invariance In harmonic superspace,''
    in *Batalin, I.A. (ed.) et al.: Quantum Field Theory and Quantum Statistics, Vol. 2*, 233-248
  and Dubna JINR - 85-363 (85,Rec.Jul.) 18p

\bibitem{Galperin:1987ek} 
  A.~S.~Galperin, E.~A.~Ivanov, V.~I.~Ogievetsky and E.~Sokatchev,
  ``$N=2$ Supergravity in Superspace: Different Versions and Matter Couplings,''
  Class.\ Quant.\ Grav.\  {\bf 4}, 1255 (1987).
 
\bibitem{Bagger:1987rc} 
  J.~A.~Bagger, A.~S.~Galperin, E.~A.~Ivanov and V.~I.~Ogievetsky,
  ``Gauging $N=2 \sigma$ Models in Harmonic Superspace,''
  Nucl.\ Phys.\ B {\bf 303}, 522 (1988).  


\bibitem{GIOS}
A.~S.~Galperin, E.~A.~Ivanov, V.~I.~Ogievetsky and E.~S.~Sokatchev,
{\it Harmonic Superspace}, Cambridge University Press, 
Cambridge, 2001.

\bibitem{Galperin:1987em} 
  A.~S.~Galperin, N.~A.~Ky and E.~Sokatchev,
  ``$N=2$ Supergravity in Superspace: Solution to the Constraints,''
  Class.\ Quant.\ Grav.\  {\bf 4}, 1235 (1987).

\bibitem{Butter:2015nza} 
  D.~Butter,
  ``On conformal supergravity and harmonic superspace,''
  JHEP {\bf 1603}, 107 (2016)
  [arXiv:1508.07718 [hep-th]].



\bibitem{KT-M_5D3} 
S.~M.~Kuzenko and G.~Tartaglino-Mazzucchelli,
  ``5D supergravity and projective superspace,''
JHEP {\bf 0802}, 004 (2008)
  [arXiv:0712.3102 [hep-th]].

\bibitem{K06} 
  S.~M.~Kuzenko,
  ``On compactified harmonic/projective superspace, 5D superconformal theories, 
  and all that,'' Nucl.\ Phys.\ B {\bf 745}, 176 (2006)
  [hep-th/0601177].
  
\bibitem{K07}
S.~M.~Kuzenko, ``On superconformal projective hypermultiplets,''
JHEP {\bf 0712}, 010 (2007) [arXiv:0710.1479].
    
  \bibitem{KLR}
A. Karlhede, U. Lindstr\"om and M. Ro\v cek,
``Self-interacting tensor multiplets in N=2 superspace,''
Phys.\ Lett.\ B {\bf 147}, 297 (1984).

\bibitem{LR1}
U.~Lindstr\"om and M.~Ro\v{c}ek,
``New hyperk\"ahler  metrics  and new supermultiplets,''
  Commun.\ Math.\ Phys.\  {\bf 115}, 21 (1988).
  
\bibitem{LR2}
U.~Lindstr\"om and M.~Ro\v{c}ek,  
 ``N=2 super Yang-Mills theory in projective superspace,''
Commun.\ Math.\ Phys.\  {\bf 128}, 191 (1990).   

\bibitem{KLRT-M_4D-1}
S.~M.~Kuzenko, U.~Lindstr\"om, M.~Ro\v{c}ek and G.~Tartaglino-Mazzucchelli,
``4D N=2 supergravity and projective superspace,'' 
JHEP {\bf 0809}, 051 (2008) [arXiv:0805.4683 [hep-th]].
 
\bibitem{KLRT-M_4D-2}
S.~M.~Kuzenko, U.~Lindstr\"om, M.~Ro\v cek and G.~Tartaglino-Mazzucchelli,
``On conformal supergravity and projective superspace,''
JHEP {\bf 0908}, 023 (2009)
[arXiv:0905.0063 [hep-th]].

\bibitem{GL} 
  J.~Grundberg and U.~Lindstr\"om,
  ``Actions for linear multiplets in six dimensions,''
  Class.\ Quant.\ Grav.\  {\bf 2}, L33 (1985).

\bibitem{BN12} 
  D.~Butter and J.~Novak,
  ``Component reduction in N=2 supergravity: 
  the vector, tensor, and vector-tensor multiplets,''
  JHEP {\bf 1205}, 115 (2012)
  [arXiv:1201.5431 [hep-th]].
  
\bibitem{Butter-hyper} 
  D.~Butter,
  ``Projective multiplets and hyperk\"ahler cones in conformal supergravity,''
  JHEP {\bf 1506}, 161 (2015)
  [arXiv:1410.3604 [hep-th]].  
  
\bibitem{BKNT-M2} 
  D.~Butter, S.~M.~Kuzenko, J.~Novak and G.~Tartaglino-Mazzucchelli,
  ``Conformal supergravity in three dimensions: Off-shell actions,''
  JHEP {\bf 1310}, 073 (2013)
  [arXiv:1306.1205 [hep-th]].
  
  
\bibitem{KNT-M} 
  S.~M.~Kuzenko, J.~Novak and G.~Tartaglino-Mazzucchelli,
  ``N=6 superconformal gravity in three dimensions from superspace,''
  JHEP {\bf 1401}, 121 (2014)
  [arXiv:1308.5552 [hep-th]].
  
\bibitem{BdeWKL} 
  D.~Butter, B.~de Wit, S.~M.~Kuzenko and I.~Lodato,
  ``New higher-derivative invariants in N=2 supergravity and the Gauss-Bonnet term,''
  JHEP {\bf 1312}, 062 (2013)
  [arXiv:1307.6546 [hep-th], arXiv:1307.6546].

\bibitem{Muller} M. M\"uller, {\it Consistent Classical Supergravity Theories},
(Lecture Notes in Physics, Vol. 336),
Springer, Berlin, 1989. 
  
\bibitem{K-08}
S.~M.~Kuzenko,
``On N = 2 supergravity and projective superspace: Dual formulations,''
Nucl.\ Phys.\  B {\bf 810}, 135 (2009)
[arXiv:0807.3381 [hep-th]].

\bibitem{KT-M-09}
 S.~M.~Kuzenko and G.~Tartaglino-Mazzucchelli,
``Different representations for the action principle in 4D N = 2 supergravity,''
  JHEP {\bf 0904}, 007 (2009)
  [arXiv:0812.3464 [hep-th]].

\bibitem{Sokatchev:PhiPrepot} 
  E.~Sokatchev,
  ``Off-shell six-dimensional supergravity in harmonic superspace,''
  Class.\ Quant.\ Grav.\  {\bf 5}, 1459 (1988).

  
\bibitem{BPB} 
  L.~Bonora, P.~Pasti and M.~Bregola,
  ``Weyl cocycles,''
  Class.\ Quant.\ Grav.\  {\bf 3}, 635 (1986).
  
\bibitem{KMM} 
  D.~R.~Karakhanian, R.~P.~Manvelyan and R.~L.~Mkrtchian,
  ``Trace anomalies and cocycles of Weyl and diffeomorphism groups,''
  Mod.\ Phys.\ Lett.\ A {\bf 11}, 409 (1996)
  [hep-th/9411068]. 
  
\bibitem{Wunsch} 
  V.~W\"unsch,
  ``Some new conformal covariants,''
  Journal of Analysis and Its Applications, {\bf 19},  
  339 (2000).
  
\bibitem{Park98} 
  J.~H.~Park,
``Superconformal symmetry in six dimensions and its reduction to four dimensions,''
Nucl.\ Phys.\ B {\bf 539}, 599 (1999)
[hep-th/9807186].

\bibitem{Siegel78} 
  W.~Siegel,
  ``Superfields in higher dimensional space-time,''
  Phys.\ Lett.\ B {\bf 80}, 220 (1979).
  
\bibitem{Nilsson} 
  B.~E.~W.~Nilsson,
  ``Superspace action for a 6-dimensional non-extended supersymmetric 
  Yang-Mills theory,'' Nucl.\ Phys.\ B {\bf 174}, 335 (1980).

\bibitem{Wess81} 
  J.~Wess,
  ``Supersymmetric gauge theories,''
  in Proceedings of the 5th Johns Hopkins Workshop on Current Problems in Particle Theory: Unified Field Theories and Beyond (25-27 May 1981, Baltimore, Maryland).
   \href{http://ccdb5fs.kek.jp/cgi-bin/img_index?198109099}{http://ccdb5fs.kek.jp/cgi-bin/img\_index?198109099}

\bibitem{HST}
P.~S.~Howe, G.~Sierra and P.~K.~Townsend,
``Supersymmetry in six dimensions,''
Nucl.\ Phys.\ B {\bf 221} (1983) 331.
  
\bibitem{KNT-M15} 
  S.~M.~Kuzenko, J.~Novak and G.~Tartaglino-Mazzucchelli,
  ``Higher derivative couplings and massive supergravity in three dimensions,''
  JHEP {\bf 1509}, 081 (2015)
  [arXiv:1506.09063 [hep-th]].
  
\bibitem{HSWest} 
  P.~S.~Howe, K.~S.~Stelle and P.~C.~West,
  ``$N=1$, $d = 6$ harmonic superspace''
  Class.\ Quant.\ Grav.\  {\bf 2}, 815 (1985).

\bibitem{Zupnik:1986da} 
  B.~M.~Zupnik,
  ``Six-dimensional Supergauge Theories in the Harmonic Superspace,''
  Sov.\ J.\ Nucl.\ Phys.\  {\bf 44}, 512 (1986)
  [Yad.\ Fiz.\  {\bf 44}, 794 (1986)].
  
\bibitem{GIKOS}
A.~S.~Galperin, E.~A.~Ivanov, S.~N.~Kalitzin, V.~Ogievetsky, E.~Sokatchev, 
``Unconstrained $N=2$ matter, Yang-Mills and supergravity theories in harmonic
superspace,''
Class.\ Quant.\ Grav.\  {\bf 1}, 469 (1984).


\bibitem{ALR14} 
  C.~Arias, W.~D.~Linch III and A.~K.~Ridgway,
  ``Superforms in simple six-dimensional superspace,''
JHEP {\bf 1605}, 016 (2016)
  [arXiv:1402.4823 [hep-th]].    
  
\bibitem{SSW}
  M.~F.~Sohnius, K.~S.~Stelle and P.~C.~West,
 ``Representations of extended supersymmetry,''
in {\it Superspace and Supergravity}, S. W. Hawking and M. Ro\v{c}ek (Eds.), 
Cambridge University Press, Cambridge, 1981, p. 283.
  
\bibitem{Hasler}
M.~F.~Hasler,
  ``The three form multiplet in N=2 superspace,''
  Eur.\ Phys.\ J.\ C {\bf 1}, 729 (1998)
  [hep-th/9606076].
  
\bibitem{Ectoplasm} 
S.~J.~Gates, Jr., ``Ectoplasm has no topology: The prelude,''
in {\it Supersymmetries and Quantum Symmetries},
 J. Wess and E. A. Ivanov (Eds.), Springer, Berlin, 1999, p. 46, arXiv:hep-th/9709104;
``Ectoplasm has no topology,''
 Nucl.\ Phys.\  B {\bf 541}, 615 (1999)
 [arXiv:hep-th/9809056].
 
\bibitem{GGKS}
S.~J.~Gates, Jr., M.~T.~Grisaru, M.~E.~Knutt-Wehlau and W.~Siegel,
``Component actions from curved superspace: Normal coordinates and
ectoplasm,'' Phys.\ Lett.\  B {\bf 421}, 203 (1998)
[hep-th/9711151].

\bibitem{Castellani} 
  L.~Castellani, R.~D'Auria and P.~Fre,
{\it Supergravity and superstrings: A Geometric perspective. Vol. 2: Supergravity},
World Scientific,  Singapore, 1991, pp. 680--684. 
    
\bibitem{HS98} 
  P.~S.~Howe and E.~Sezgin,
  ``Anomaly free tensor Yang-Mills system and its dual formulation,''
  Phys.\ Lett.\ B {\bf 440}, 50 (1998)
  [hep-th/9806050].
  
\bibitem{KNS15} 
  S.~M.~Kuzenko, J.~Novak and I.~B.~Samsonov,
  ``The anomalous current multiplet in 6D minimal supersymmetry,''
  JHEP {\bf 1602}, 132 (2016)
  [arXiv:1511.06582 [hep-th]].
  
\bibitem{GKT-M09} 
  S.~J.~Gates, Jr., S.~M.~Kuzenko and G.~Tartaglino-Mazzucchelli,
  ``Chiral supergravity actions and superforms,''
  Phys.\ Rev.\ D {\bf 80}, 125015 (2009)
  [arXiv:0909.3918 [hep-th]].
  
  \bibitem{Muller89}
M.~M\"uller, ``Off-shell supergravity actions,''
Preprint MPI-PAE/PTh 64/89, Munich,  1989.

\bibitem{BKN:LM}
  D.~Butter, S.~M.~Kuzenko and J.~Novak,
  ``The linear multiplet and ectoplasm,''
  JHEP {\bf 1209} (2012) 131
  [arXiv:1205.6981 [hep-th]].

 \bibitem{KN14:5DCS} 
  S.~M.~Kuzenko and J.~Novak,
  ``On supersymmetric Chern-Simons-type theories in five dimensions,''
  JHEP {\bf 1402}, 096 (2014)
  [arXiv:1309.6803 [hep-th]].

\bibitem{KN:3DCS}
  S.~M.~Kuzenko and J.~Novak,
  ``Supergravity-matter actions in three dimensions and Chern-Simons terms,''
  JHEP {\bf 1405} (2014) 093
  [arXiv:1401.2307 [hep-th]].

\bibitem{Butter:Proj}
  D.~Butter,
  ``New approach to curved projective superspace,''
  Phys.\ Rev.\ D {\bf 92} (2015) no.8,  085004
  [arXiv:1406.6235 [hep-th]].

\bibitem{Butter:Harm}
  D.~Butter,
  ``On conformal supergravity and harmonic superspace,''
  JHEP {\bf 1603} (2016) 107
  [arXiv:1508.07718 [hep-th]].
  
  \bibitem{Novak:VT1} 
  J.~Novak,
  ``Superform formulation for vector-tensor multiplets in conformal supergravity,''
  JHEP {\bf 1209}, 060 (2012)
  [arXiv:1205.6881 [hep-th]].
  
  \bibitem{Novak:VT2} 
  J.~Novak,
  ``Variant vector-tensor multiplets in supergravity: Classification and component reduction,''
  JHEP {\bf 1303}, 053 (2013)
  [arXiv:1210.8325 [hep-th]].

\bibitem{BT15} 
  M.~Beccaria and A.~A.~Tseytlin,
  ``Conformal anomaly c-coefficients of superconformal 6d theories,''
  JHEP {\bf 1601}, 001 (2016)
  [arXiv:1510.02685 [hep-th]].
  
\bibitem{BNT-M} 
Daniel Butter, Joseph Novak and Gabriele Tartaglino-Mazzucchelli, \emph{to appear}.



  
  


\end{thebibliography}
\end{document}